\documentclass[times,final]{elsarticle}

\usepackage{jcomp}
\usepackage{framed,multirow}

\usepackage{amssymb}
\usepackage{latexsym}

\usepackage{url}
\usepackage{mathtools}
\usepackage{xcolor}
\usepackage{placeins}
\definecolor{newcolor}{rgb}{.8,.349,.1}
\usepackage[outdir=./]{epstopdf}
\usepackage{bm}
\usepackage{longtable}
\usepackage{romannum}

\journal{Journal of Computational Physics}
\begin{document}
\pagenumbering{arabic}
\verso{Bhargav Mantravadi \textit{etal}}
\begin{frontmatter}

\title{A hybrid discrete exterior calculus and finite difference method for Boussinesq convection in spherical shells}%

\author{Bhargav \snm{Mantravadi}}
\author{Pankaj \snm{Jagad}\corref{cor1}}
\ead{pankaj.jagad@kaust.edu.sa}
\author{Ravi \snm{Samtaney}}
\cortext[cor1]{Corresponding author:}

\address{Mechanical Engineering, Division of Physical Science and Engineering, King Abdullah University of Science and Technology, Thuwal, Saudi Arabia}

\begin{abstract}
We present a new hybrid discrete exterior calculus (DEC) and finite difference (FD) method to simulate fully three-dimensional Boussinesq convection in spherical shells subject to internal heating and basal heating, relevant in the planetary and stellar phenomenon. We employ DEC to compute the surface spherical flows, taking advantage of its unique features of structure preservation (e.g., conservation of secondary quantities like kinetic energy) and coordinate system independence, while we discretize the radial direction using FD method. The grid employed for this novel method is free of problems like the coordinate singularity, grid non-convergence near the poles, and the overlap regions. We have developed a parallel in-house code using the PETSc framework to verify the hybrid DEC-FD formulation and demonstrate convergence. We have performed a series of numerical tests which include quantification of the critical Rayleigh numbers for spherical shells characterized by aspect ratios ranging from 0.2 to 0.8, simulation of robust convective patterns in addition to stationary giant spiral roll covering all the spherical surface in moderately thin shells near the weakly nonlinear regime, and the quantification of Nusselt and Reynolds numbers for basally heated spherical shells. \textcolor{black}{The method exhibits slightly better than second order error convergence with the mesh size.}
\end{abstract}



\end{frontmatter}


\section{Introduction}
\label{introduction}

Solar convection is a complex phenomenon due to the following reasons. First, the buoyant flow is highly turbulent, where the relevant non-dimensional parameters have extreme values. The Rayleigh number ($Ra$), which characterizes the relative importance of buoyancy and viscous forces, is in the range of  $10^{21} - 10^{24}$; the Prandtl number ($Pr$), which is the ratio of momentum diffusivity and thermal diffusivity, is of the order of $10^{-7} -10^{-4}$ \cite{hanasoge2016seismic}. Second, the Sun is subject to differential rotation. The period of rotation at the equator is 25 days while that of the polar regions is 33 days \cite{miesch2005large}. Additionally, the fluid flow is affected by the influence of Coriolis force, and the Taylor number, which characterizes the relative importance of rotational and viscous forces, and is of the order of $10^{21} - 10^{29}$ \cite{garcia2018onset}. Finally, there is self-generation, and transport of magnetic fields due to the convective plasma motion \cite{fan2021magnetic}; see also \cite{miesch2005large,fan2021magnetic,spiegel1971convection,hughes1988magnetic,spruit1990solar,miesch2009turbulence,rieutord2010sun,kupka2017modelling}. Although the complexity of the problem can be reduced significantly by making the Boussinesq approximation, it is still a major challenge to simulate this phenomenon, given the extreme parameter regime and the complex geometry. Furthermore, Boussinesq equations are of interest in themselves as they provide a basis for a better understanding of solar convection. We thus study this classical fluid dynamical problem of Boussinesq convection in spherical shells with a major impetus of solar convection. 

The literature reporting convection studies in spherical geometry are less numerous in comparison to that in the planar and the cylindrical geometries \cite{gastine2015turbulent,gastine2016scaling}. Further, it is debatable if the results of the planar and cylindrical geometries are directly applicable to the spherical geometry. Through massively parallel simulations of convection in the slender cylindrical cells, Iyer et al. \cite{iyer2020classical} pushed the Rayleigh number up to $10^{15}$ and observed the turbulent heat transport to continue to follow the classical $1/3$ scaling law. Nevertheless, in the case of spherical shells, the presence of `Ultimate Regime' ($Ra\geq10^{12}$, $\text{Nusselt number}$ $Nu \sim Ra^{1/2}$), characterized by enhanced heat transfer, is still unclear. Based on the Rayleigh-B\'enard convection (RBC) study in spherical shells for Rayleigh numbers up to $10^9$, the suggestion of Gastine et al. \cite{gastine2015turbulent} of an early transition to the enhanced heat transfer regime, at a lower $Ra$ compared to that in a planar geometry, is yet to be established. Besides, the effect of basal and internal heating on the scaling laws at extreme Rayleigh numbers remains to be verified. To that end, pushing the current limits to explore an extreme parameter regime, especially in the case of spherical geometry, requires a considerable effort in the high-performance computing front and beseeches for possible alternate numerical methods.

Recent years have witnessed an increasing usage of numerical methods that preserve the mathematical and physical properties of the governing equations, which are usually partial differential equations, by employing the language of ``Differential forms". Such numerical methods have an associated exact Discrete calculus, where the calculus and physics are discretized exactly, and all the approximation occurs in the material constitutive equations \cite{perot2007discrete}; the orthogonality relations $\nabla \cdot \nabla \times \bm{A} = 0$, for any vector $\bm{A}$, $\nabla \times \nabla a = 0$, for any scalar $a$, are respected discretely. The calculus of differential forms is called Exterior calculus, and its extension to discrete spaces is called the Discrete Exterior Calculus (DEC). \textcolor{black}{DEC discretely mimics the Stokes and the divergence theorem by the usage of a dual mesh, in addition to the primal simplicial mesh}. DEC framework employed in this work is based on the approach put forward by Hirani et al. \cite{hirani2003discrete,desbrun2005discrete}. Although DEC was initially limited to computer graphics and computer vision \cite{desbrun2003discrete,desbrun2008discrete}, in recent years it has been finding applications in the field theories like electromagnetism \cite{chen2016discrete} and fluid dynamics \cite{hirani2015numerical,mohamed2016discrete,jagad2020investigation,jagad2021primitive,jagad2021effects,wang2022discrete}; see Mohamed et al. \cite{mohamed2018numerical} and Schulz and Tsogtgerel \cite{schulz2020convergence} for numerical and analytical evidence, respectively, for the first-order convergence in DEC. 


Taking advantage of these unique features of coordinate independent nature, and structure preservation properties of DEC, we embark on developing a hybrid discrete exterior calculus and finite difference (DEC-FD) method to simulate convection in spherical shells. The novelty resides in deriving the operator-split governing equations, followed by replacing the surface spherical operators with the DEC operators and approximating the radial operators using the FD operators. This hybrid method preserves solutions that are only radially dependent and gives control on the mesh refinement in the surface and radial direction independently. Further, the grid used in this method is free of problems like coordinate singularity, grid non-convergence near the poles, and overlap regions, which are often encountered in the cubed-sphere and Yin-Yang grids. The objectives of the present work are twofold: (1) to present the DEC-FD methodology for convection simulation in spherical shells; (2) to verify the hybrid formulation using a series of numerical experiments. Furthermore, we only intend to demonstrate the possibility and provide the necessary ingredients to combine DEC and FD for simulating convection in spherical shells. The chosen spatial finite difference scheme, time integration scheme, linear solver, and parallelization may not be the optimal choice. 

The remainder of this paper is organized as follows. In Section~\ref{section: governing equations}, we present the non-dimensional Oberbeck-Boussinesq equations and derive the operator-split equations in vector calculus notation, which are further transformed into exterior calculus notation. Section~\ref{section: DEC-FD hybrid method} discusses the discretization, reviews DEC and its corresponding important operators - the discrete exterior derivative, the Hodge star, and the wedge product, and presents generalized DEC-FD equations. In Section~\ref{section: Numerical Experiments}, we demonstrate the convergence and verify the DEC-FD formulation by comparing the numerical results with theory and literature. Finally, in Section~\ref{section: Conclusion}, conclusions are drawn, highlighting the directions for future development. 

\section{Governing equations}
\label{section: governing equations}

\textcolor{black}{We first write the governing continuity, momentum, and energy equations in vector calculus notation in three-dimensions. Later, we split the operators and equations in surface-spherical and radial directions. Subsequently, we express the equations in exterior calculus notation.}

We consider convection, under Boussinesq approximation, of a fluid with momentum diffusivity $\nu$, thermal diffusivity $\kappa$, and thermal expansivity $\alpha$, confined in a  spherical shell of inner radius $R_0$ and outer radius $R_1$, subject to a spherically symmetric gravitational field $\bm{g} = - g(r) \ \bm{e_r}$ and an unstable temperature gradient $\nabla T_c(r) = -f(r)\  \bm{e_r}$. Here, $r$ is the radius, and $\bm{e_r}$ is the unit vector in the radial direction. Using $l_o$ as the length scale, $t_o$ as the time scale, $\theta_o'$ as the unit of temperature fluctuation, we obtain the non-dimensional Oberbeck-Boussinesq equations expressed by

\begin{equation}
\begin{split}
\nabla \cdot \bm{u} = 0 ,
\end{split}
\label{eq: non-dimensional continuity}
\end{equation}

\begin{equation}
\frac{\partial \bm{u}}{\partial t} + \bm{u}  \cdot  \nabla  \bm{u}  = - \nabla p + \zeta(r) \   \theta' \bm{e_r} + \nabla^2 \bm{u},
\label{eq: non-dimensional momentum equation}
\end{equation}

\begin{equation}
\psi \left( \frac{\partial \theta'}{\partial t} + \bm{u} \cdot \nabla   \theta' \right)  = \chi(r) \ u_r + \xi \nabla^2 \theta',
\label{eq: non-dimensional energy equation}
\end{equation}

%
%

\noindent where $t$, $\bm{u}$, $p$, and $\theta'$ are the time, the velocity vector field, the pressure, and the temperature fluctuation from base state temperature $T_c(r)$, respectively. The coefficients $\zeta(r)$, $\psi$, $\chi(r)$, and $\xi$ depend on the functions $g(r)$ and $f(r)$ and \textcolor{black}{various reference scales employed for} the non-dimensionalization, which will be discussed later.

We then split the differential operators as surface and radial operators. For instance, $\nabla  = \nabla_\perp + \bm{e_r} \frac{\partial}{\partial r}$, where the subscript $\perp$ denotes the spherical surface. Similarly, we split the three-dimensional velocity vector field $\bm{u}  = \bm{u}_\perp + u_r \bm{e_r}$, where the spherical surface velocity vector $\bm{u}_\perp = u_\theta \bm{e_\theta} + u_\phi \bm{e_\phi}$,  $\bm{e_\theta}$ is the unit vector in the colatitudinal direction $\theta$, $\bm{e_\phi}$ is the unit vector in the longitudinal direction $\phi$, and $u_r$ is the velocity in the radial direction with unit vector $\bm{e_r}$. Employing operator splitting and vector field splitting in the foregoing Eqs. (\ref{eq: non-dimensional continuity}) $-$ (\ref{eq: non-dimensional energy equation}), we \textcolor{black}{have}

\begin{equation}
\begin{split}
\nabla_\perp \cdot \bm{u}_\perp + \frac{1}{r^2}\frac{\partial}{\partial r} \left( r^2 u_r \right) = 0,
\end{split}
\label{eq: operator-split continuity equation}
\end{equation}

\begin{equation}
\begin{split}
\frac{\partial \bm{u}_\perp}{\partial t} + \bm{u}_\perp \cdot \nabla_\perp \bm{u}_\perp  + \frac{u_r}{r}\bm{u}_\perp + u_r \frac{\partial \bm{u}_\perp}{\partial r} = - \nabla_\perp p + 
\left(  \nabla_\perp \left(\nabla_\perp \cdot \bm{u}_\perp \right)
-\Delta^R_\perp  \bm{u}_\perp + \frac{2 \bm{u}_\perp}{r^2} \color{black} + \frac{2}{r} \nabla_\perp u_r + \frac{2}{r} \frac{\partial \bm{u}_\perp}{\partial r} + \frac{\partial^2 \bm{u}_\perp}{\partial r^2}\right),
\end{split}
\label{eq: operator-split surface momentum balance}
\end{equation}

\begin{equation}
\begin{split}
\frac{\partial u_r}{\partial t} + \bm{u}_\perp \cdot \nabla_\perp u_r - \frac{|\bm{u}_\perp|^2}{r} + u_r \frac{\partial u_r}{\partial r} = - \frac{\partial p}{\partial r} + \zeta(r) \ \theta'  + \left( \nabla_\perp^2  u_r - \frac{4}{r} \nabla_\perp \cdot \bm{u}_\perp - \frac{ 6 u_r }{r^2} +\frac{\partial^2 u_r}{\partial r^2} \right),
\end{split}
\label{eq: operator-split radial momentum balance}
\end{equation}

\begin{equation}
\begin{split}
\psi \left( 
\frac{\partial \theta'}{\partial t} 
+ \bm{u}_\perp \cdot \nabla_\perp \theta' 
+ u_r \frac{\partial \theta'}{\partial r}
\right) = \chi(r) \ u_r + \xi \left(  \nabla_\perp^2  \theta' +  \frac{\partial^2  \theta'}{\partial r^2} + \frac{2}{r} \frac{\partial  \theta'}{\partial r} \right),
\end{split}
\label{eq: operator-split energy balance}
\end{equation}
\noindent where Eqs. (\ref{eq: operator-split continuity equation}) $-$ (\ref{eq: operator-split energy balance}) denote the continuity, surface momentum balance, radial momentum balance, and energy balance, respectively. The surface Laplace–DeRham operator is denoted by $\Delta^R_\perp$. The operator $\Delta^R_\perp$ and the Gaussian curvature term ${2 \bm{u}_\perp}/{r^2}$ in the Eq. (\ref{eq: operator-split surface momentum balance}) are a result of the divergence of the deformation tensor and the non-commutativity of the second covariant derivative in curved spaces; see Aris \cite[pp. 235]{aris2012vectors} and Nitschke et al. \cite{nitschke2017discrete}. By annihilating $u_r$ and the terms with radial dependence, we recover the equations governing the two-dimensional flow on the spherical surface \cite{nitschke2017discrete}. 

 We express the preceding Eqs. (\ref{eq: operator-split continuity equation}) $-$ (\ref{eq: operator-split energy balance}) in the exterior calculus notation before deriving them in the DEC notation \textcolor{black}{(\textcolor{black}{we recommend for the reader unfamiliar with the topic of exterior calculus to read through} Section \ref{subsection: DEC} and References \cite{mohamed2016discrete}, and \cite{jagad2021primitive} for further details, and a primer on exterior calculus and DEC)}. For this purpose, we substitute the vector calculus identities 

\begin{equation}
\begin{aligned}
\left ( \bm{u}_\perp \cdot \nabla_\perp \right) \bm{u}_\perp & = \frac{1}{2} \nabla_\perp \left ( \bm{u}_\perp \cdot \bm{u}_\perp \right ) - \bm{u}_\perp \times \left (\nabla_\perp \times \bm{u}_\perp \right ) , \\
\left ( \bm{u}_\perp \cdot \nabla_\perp \right) u_r & = \nabla_\perp \cdot (u_r \bm{u}_\perp) - u_r  (\nabla_\perp \cdot \bm{u}_\perp) ,\\
\end{aligned}
\label{eq: vector calculus identities}
\end{equation}

\noindent in Eqs. (\ref{eq: operator-split continuity equation})  $-$ (\ref{eq: operator-split energy balance}), and apply flat operator $\flat$ to carry out the notation transformation from vector calculus to exterior calculus by employing the identities

\begin{equation}
\begin{aligned}
\left( a \bm{k}  \right)^\flat & = a \wedge \bm{k}^\flat  , 
\\
\left( a b \right)^\flat & = a \wedge b   ,
\\
\left( \nabla \cdot \bm{k}  \right)^\flat & = \ast\text{d}\ast \bm{k}^\flat  , 
\\
\left( \nabla a \right)^\flat & = \text{d} a
\\
\left( \Delta^R_\perp \bm{u}_\perp \right) ^\flat & = \left( \nabla_\perp \times \nabla_\perp \times \bm{u}_\perp \right)^\flat = \left ( -1 \right )^{N+1}  \ast\text{d}\ast \text{d}\bm{u}_\perp^\flat  , 
\\
\left[ \bm{u}_\perp \times \left (\nabla_\perp \times \bm{u}_\perp \right ) \right ]^\flat & = (-1)^{N+1} \ast \left( \bm{u}_\perp^\flat \wedge \ast \text{d} \bm{u}_\perp^\flat \right)  , 
\\
\end{aligned}
\label{eq: exterior calculus identities}
\end{equation}


\noindent \textcolor{black}{where $a$ and $b$ are scalars, $\bm{k}$ is a vector, $\bm{k}^\flat$ and $\bm{u}_\perp^\flat$ are 1-forms, $\ast$ is the Hodge star, $\wedge$ is the wedge product, $\text{d}$ is the exterior derivative operator,} $N$ is the space dimension, and here $N = 2$ as we apply exterior calculus only to spherical surfaces. Substituting Eqs. (\ref{eq: vector calculus identities}) and (\ref{eq: exterior calculus identities}) in Eqs. (\ref{eq: operator-split continuity equation}) $-$ (\ref{eq: operator-split energy balance}), we obtain the equations in the exterior calculus notation
\begin{equation}
\begin{split}
\ast \ \text{d}  \ast \bm{u}_\perp^\flat  
+  \frac{1}{r^2}\frac{\partial}{\partial r} \left( r^2 u_{r} \right)  = 0 ,
\end{split}
\label{eq: split governing continuity equation}
\end{equation}

\begin{equation}
\begin{split}
\frac{\partial \bm{u}_\perp^\flat }{\partial t} 
+ \ast \ \bm{u}_\perp^\flat  \wedge \ast \text{d} \bm{u}_\perp^\flat  
+ \frac{u_r}{r}\wedge \bm{u}_\perp^\flat 
+ u_r \wedge \frac{\partial \bm{u}_\perp^\flat }{\partial r}  
+ \text{d} p 
+ \frac{1}{2} \text{d} \left( \bm{u}_\perp \cdot \bm{u}_\perp \right )^\flat
- \left(  \text{d} \ast \text{d} \ast  \bm{u}_\perp^\flat
+ \ast \ \text{d} \ast \text{d}  \bm{u}_\perp^\flat
+ \frac{2 \bm{u}_\perp^\flat}{r^2} 
+ \frac{2}{r} \text{d} u_r 
+
\right .
\\
\left .
 \frac{2}{r} \frac{\partial  \bm{u}_\perp^\flat}{\partial r}  
+ \frac{\partial^2  \bm{u}_\perp^\flat}{\partial r^2} 
\right ) = 0,
\end{split}
\end{equation}

\begin{equation}
\begin{split}
\frac{\partial u_r}{\partial t} 
+ \ast \ \text{d} \ast \left( \bm{u}_\perp^\flat \wedge u_r \right) 
- u_r \wedge \left( \ast \ \text{d} \ast \bm{u}_\perp^\flat \right) 
- \frac{\left( \bm{u}_\perp \cdot \bm{u}_\perp \right)^\flat}{r}
+ u_r \wedge \frac{\partial u_r}{\partial r}  
+ \frac{\partial p}{\partial r} 
- \zeta(r) \  \theta'  
-  \left(  \ast \ \text{d} \ast \text{d} u_r 
- \frac{4}{r} \left( \ast \ \text{d}  \ast \bm{u}_\perp^\flat \right) 
- 
\right .
\\
\left.
\frac{ 6 u_r }{r^2} 
+\frac{\partial^2 u_r}{\partial r^2} 
\right ) = 0 ,
\end{split}
\end{equation}

\begin{equation}
\begin{split}
\psi \left ( \frac{\partial \theta'}{\partial t} 
+ \ast\  \text{d} \ast \left( \bm{u}_\perp^\flat \wedge \theta' \right) 
- \theta' \wedge \left( \ast \ \text{d} \ast \bm{u}_\perp^\flat \right)  
+ u_r \wedge \frac{\partial \theta'}{\partial r} \right) 
- \chi(r) \  u_r 
- \xi \left( \ast \text{d} \ast \text{d} \ \theta' 
+  \frac{\partial^2 \theta'}{\partial r^2} 
+ \frac{2}{r} \frac{\partial \theta'}{\partial r} 
\right ) = 0 ,
\end{split}
\label{eq: split governing energy equation}
\end{equation}

\noindent where the flat operator $\flat$ on the vector $\bm{u}_\perp$ maps onto a 1-form $\bm{u}^\flat_\perp$. Similarly, pressure, radial velocity, and temperature fluctuations are transformed into 0-forms $p$, $u_r$, and $\theta'$, respectively.

\section{DEC-FD hybrid methodology}
\label{section: DEC-FD hybrid method}

We shall now move to discrete space. In this section, we describe the domain discretization, give a brief introduction to DEC \textcolor{black}{for the purposes of this work}, and express the equations in the DEC-FD notation, where the smooth differential forms, exterior calculus operators, and radial differential operators are now replaced with discrete forms, discrete exterior calculus operators, and finite difference operators, respectively.

\subsection{Domain discretization}
\label{subsection: Domain discretization}

\begin{figure}[ht]
\centering
\begin{center}
\begin{tabular}{c c}
\includegraphics[scale=0.23]{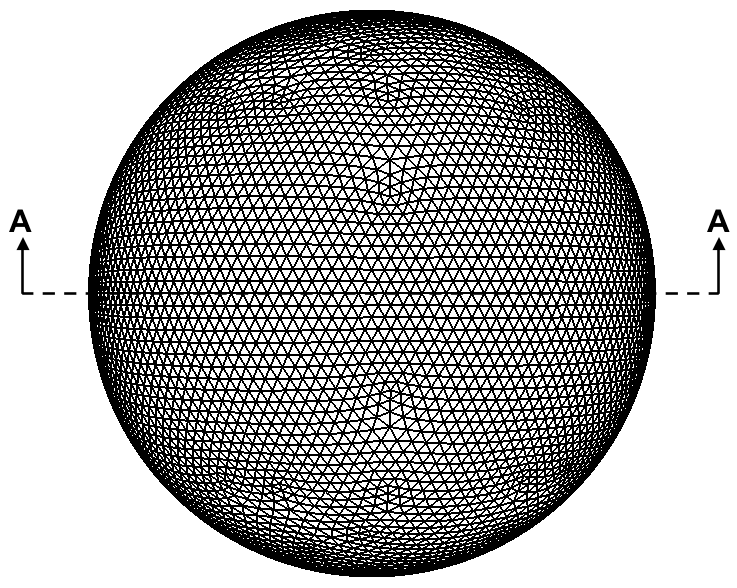} &
\includegraphics[scale=0.17]{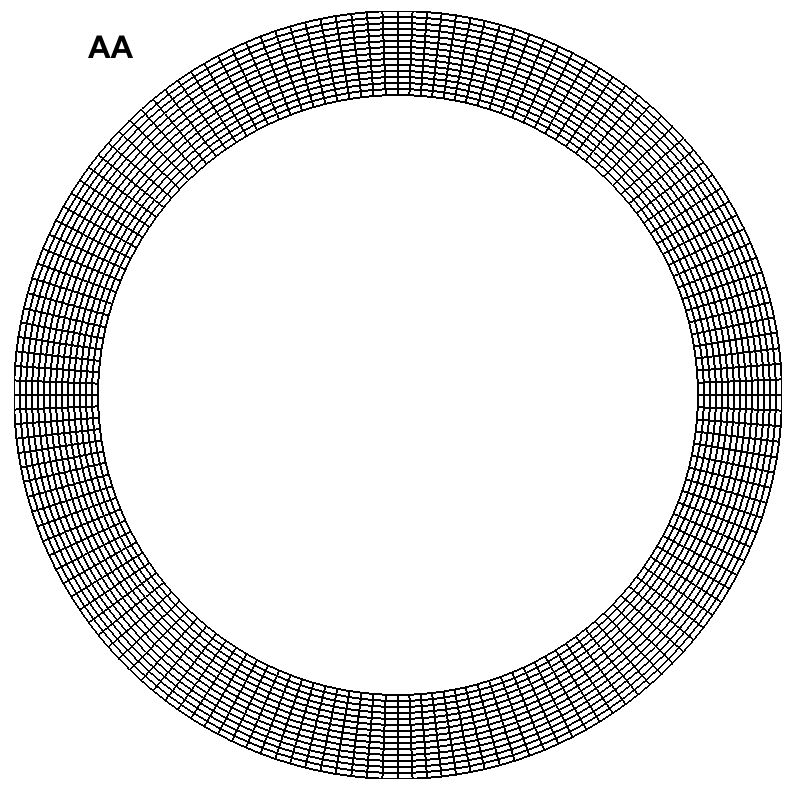} \\
\end{tabular}
\end{center}
\caption{Prismatic mesh showing triangulation of the outer most spherical layer along with the meridional sectional view.}
\label{fig: Prismatic mesh}
\end{figure}

We first obtain a base spherical surface mesh with the radius of the inner sphere $R_0$ by recursive construction of the icosahedral grid, with grid refinement level $s_r$; see Wang and Lee \cite{wang2011geometric} for details of the geometric properties of icosahedral grids. The number of triangular elements on the base surface spherical mesh is \textcolor{black}{given by} $20 q^2$, where $q = 2^{s_r}$. The prismatic mesh, as shown in Fig. \ref{fig: Prismatic mesh}, is obtained by extruding the base surface mesh in the radial direction. The prismatic mesh, however, is not explicitly stored. The resolution in the radial direction is controlled by the number of spherical surface layers $k$, excluding the innermost and outermost boundary spheres. The final mesh can also be thought of as the stacking of $k$ spherical layers in the radial direction \textcolor{black}{without the boundary spheres.} The spherical layers are denoted by subscript $i$; $i$ varying from $0$ to $k-1$.

Let $\Omega_i$ be the spherical surface domain of dimension $N = 2$ for the $i^{\text{th}}$ spherical layer. \textcolor{black}{From here on, subscript $i$ denotes $i^{\text{th}}$ spherical layer}. The domain $\Omega_i$ is approximated by simplicial complex $K_i$; see Munkres \cite{munkres1984elements} for details on simplices and simplicial complexes. The domain simplex $\sigma_i$ of dimension $j$ is denoted by $\sigma_i^j \in K_i$; see \cite{hirani2003discrete,desbrun2005discrete}. Corresponding to the primal simplicial complex $K_i$, there exists a dual complex $\star K_i$. \textcolor{black}{Further,} $(N-j)$-cell represented by $\star \sigma_i^j \in K_i$ is the corresponding dual for primal $j$-simplex $\sigma_i^j \in K_i$. We consider circumcenteric dual mesh here. For instance, the dual of a primal edge is the dual edge connecting the circumcenters of the triangles supported by the primal edge. The dual of a primal node is the polygon formed by the dual edges associated with the corresponding primal edges connected to this primal node. The dual of a triangle is its circumcenter. Note that for the case of a primal mesh representing a curved surface, the dual cells and dual edges can be non-planar. Let $\star K_i$ consist of $N_0$ dual nodes, $N_1$ dual edges, and $N_2$ dual cells. The dual mesh is not explicitly stored,  \textcolor{black}{however.}

Both primal and dual meshes are oriented. The highest dimensional simplices of the primal mesh are oriented consistently, either clockwise or counterclockwise. The lower dimensional simplices can be oriented arbitrarily. The orientation of dual cells is established based on the orientation of the primal simplices; see Hirani et al. \cite{hirani2015numerical}. Note that the orientation of the simplices of the primal mesh and their corresponding dual cells remains the same for all the $k$ spherical layers. Also, the simplicial meshes considered here are Delaunay meshes. 

\FloatBarrier
\FloatBarrier

\subsection{Discrete exterior calculus (DEC)}
\label{subsection: DEC}
We shall now give a brief review of DEC with restriction to two dimensions. More information on DEC can be found in \cite{hirani2003discrete,desbrun2005discrete,mohamed2016discrete,hirani2015numerical,crane2018discrete}. Exterior calculus is the calculus of ``Differential forms." The Differential forms, as described by Flanders \cite{flanders1963differential}, are ``the things which occur under the integral signs." For instance, consider a line integral of a vector $\bm{w}$, $\int \bm{w} \cdot  \bm{\text{d}l}$, where $\bm{w} \cdot  \bm{\text{d}l}$ represents a 1-form $w^1$. Note that the superscript on the form denotes the dimension of the form. Similarly, consider a surface integral, $\int \bm{p} \cdot  \bm{\text{d}A}$, where $\bm{p} \cdot  \bm{\text{d}A}$ denotes a 2-form $p^2$. Since the 0-integral is trivial, 0-form $m^0$ is simply a scalar function; see Frankel \cite{frankel2011geometry} for more information on differential forms. 


The discrete forms, on the other hand, are the entire integral quantities. \textcolor{black}{Unlike the continuous differential forms which are defined at every point in the domain, discrete forms are defined on a mesh. }A discrete differential $j$-form relates a scalar with a discrete $j$-dimensional mesh object. For instance, the discretization of the smooth velocity 1-form $\bm{u}^\flat$ is defined on the primal edges $\sigma^1_i$ (recall that subscript $i$ denotes $i^\text{th}$ spherical layer) as $\int_{\sigma_i^1} \bm{u}.\text{d}l $, or defined on the dual edges $\ast \sigma^1_i$ as $\int_{\ast\sigma_i^1} \bm{u}.\text{d}l$, denoting a primal or dual discrete1-form. The discrete 0-forms are defined as scalar quantities on the dual or primal nodes. \textcolor{black}{Similarly, the discrete 2-forms, which are the cell average integrals of the scalar field (in two-dimensions), are defined on the primal triangles or dual cells. Using discrete forms as the primary unknowns allows for exact discretization, where all the errors are restricted to the algebraic level \cite{perot2007discrete}.}

The space of discrete $j$-forms defined on the primal and dual mesh complexes are related by discrete exterior derivative and Hodge star operator. \textcolor{black}{The discrete exterior derivative operator, $\text{d}$, behaves like a multi-dimensional differentiation and when applied to a form, it takes it to the next higher dimensional form. }The operator, $\text{d}_j$, maps a primal $j$-form to a primal $\left( j+1 \right)$-form. In the case of dual complex, the transpose of $\text{d}_{\left( N-j-1 \right)}$ maps a dual $j$-form to a dual $\left( j+1 \right)$-form, with an exception of a negative sign for $\text{d}_0^T$ operator due to \textcolor{black}{convention} of the mesh orientation in two-dimensions. \textcolor{black}{For instance, in 2D, the gradient of a scalar $\nabla f = \bm{g}$ in vector calculus notation is equivalent to $\text{d}_0 f = g^1$ in DEC notation, where the vector $\bm{g}$ is best represented by a 1-form $g^1$. The operation equivalent to divergence of a vector $\nabla \cdot \bm{w} = f$ is $\text{d}_1 w^1 = f^2$, resulting in a 2-form which is a scalar quantity.}

 \textcolor{black}{The Hodge star operator takes values on the primal mesh and transfers them onto the dual mesh along with changing their dimensions. }The operator, $\ast_j$, maps a primal $j$-form to a dual $\left(N- j\right)$-form. While, the inverse Hodge star operator $\ast_j^{-1}$ maps a dual $\left(N- j\right)$-form to primal $j$-form. We consider circumcentric dual mesh, which results in the diagonal definitions for all the Hodge star operators and their inverses. \textcolor{black}{ The $n^{\text{th}}$ diagonal element of the matrix is the ratio of the volume of the dual $\left(N- j\right)$-cell $\star \sigma_{n,i}^j$ and the volume of the primal $j$-simplex $\sigma_{n,i}^j$, i.e., $\frac{\star \sigma_{n,i}^j}{\sigma_{n,i}^j}$.}  \textcolor{black}{In 2D, Hodge star operator on a 0-form (scalar point value on primal node) results in a 2-form (dual cell average of scalar) and vice versa. Hodge star when applied on a 1-form on primal edge (vector component along the line) results in a 1-form on dual edges.}

\textcolor{black}{The wedge product or exterior product, $\wedge$, between forms is simply a multiplication.} The wedge product of a discrete $j$-form $p^j$ and $l$-form $q^l$ results in a $\left( j+l \right)$-form $r^{j+l}$, i.e., $p^j \wedge q^l = r^{j+l}$, for $j+l \leq N$, with a property $p^j \wedge q^l = (-1)^{jl} q^l \wedge p^j $. \textcolor{black}{In 2D, a wedge product is either a regular multiplication or a 2D cross product.}


\subsection{DEC-FD equations}
\label{subsection: DEC}

In this work, the velocity 1-form $\bm{u}_\perp^\flat$ in all terms (except in one term to be discussed later) is discretized as a dual 1-form. This choice requires the radial velocity $u_r$, pressure $p$, and temperature fluctuations $\theta'$ to be defined as dual 0-forms. Let vector $U_{\perp,i}$ with $N_1$ entries contain the dual velocity 1-forms that are stored on the dual edges of the $i^{\text{th}}$ spherical layer. Let vectors $U_{r,i}$, $P_i$, and $\Theta_i$ with $N_0$ entries contain the dual 0-forms $u_r$, $p$, and $\theta'$, respectively, that are stored on the dual nodes of the $i^{\text{th}}$ spherical layer. We replace the smooth Exterior calculus and differential operators in Eqs. (\ref{eq: split governing continuity equation}) $-$ (\ref{eq: split governing energy equation}) by their discrete counterparts DEC and FD operators, respectively. The DEC-FD equations are

\begin{equation}
\begin{split}
 \ast_{2,i} \ \text{d}_1  \ast_1^{-1} \color{black} U_{\perp,i} +  \frac{1}{r_i^2} D_{r,1} \color{black} (r_i^2 U_{r,i})  = 0,
\end{split}
\label{eq: wedge continuity}
\end{equation}

\begin{equation}
\begin{split}
\frac{\partial U_{\perp,i}}{\partial t} 
+  \ast_1V_{\perp,i}  \ \wedge\ \ast_{0,i}^{-1} [-\text{d}_0^{\text{T}}] \ U_{\perp,i} 
+ \frac{U_{r,i}}{r_i} \wedge U_{\perp,i} 
+ U_{r,i}  \wedge   D_{r,1} \color{black}  U_{\perp,i}  \ 
+  \text{d}_1^{\text{T}} P_i \ 
+ \frac{1}{2}  \text{d}_1^{\text{T}} K_i \  
- 
 \left[ \left( \text{d}_1^T \ast_{2,i} \ \text{d}_1  \ast_1^{-1}  + \right. \right. \\
 \left. \ast_1 \ \text{d}_0 \ast_{0,i}^{-1} [ -\text{d}_0^{\text{T}} ] \right) U_{\perp,i}  
 + \frac{2 \ U_{\perp,i}}{r_i^2} 
 + \frac{2}{r_i} \  \text{d}_1^{\text{T}} \color{black} U_{r,i} 
 + \frac{2}{r_i} D_{r,1} \color{black}  U_{\perp,i} 
+ D_{r,2} U_{\perp,i} \Bigl] = 0,
\end{split}
\label{eq: wedge surface momentum balance}
\end{equation}

\begin{equation}
\begin{split}
\frac{\partial U_{r,i}}{\partial t} 
+ \ast_{2,i} \ \text{d}_1 \ast_1^{-1} (U_{\perp,i} \wedge U_{r,i}) 
- U_{r,i} \wedge(\ast_{2,i} \ \text{d}_1 \ast_1^{-1} U_{\perp,i}) 
- \frac{K_i}{r_i} 
+ U_{r,i} \wedge D_{r,1} U_{r,i} 
+  D_{r,1} P_i 
- \zeta_i \ \Theta_i 
- 
\\
 \left(  
 \ast_{2,i} \ \text{d}_1 \ast_1^{-1} \text{d}_1^T \ U_{r,i} 
 - \frac{4}{r_i} \left( \ast_{2,i} \ \text{d}_1  \ast_1^{-1} \right) U_{\perp,i}  
 - \frac{ 6 \ U_{r,i} }{r_i^2} 
 + D_{r,2} U_{r,i} 
\right ) = 0  ,
\end{split}
\label{eq: wedge radial momentum balance}
\end{equation}

\begin{equation}
\begin{split}
\psi \left( \frac{\partial \Theta_i}{\partial t} 
+ \ast_{2,i} \ \text{d}_1 \ast_1^{-1} \left( U_{\perp,i} \wedge \Theta_i \right)
- \Theta_i \wedge \left(\ast_{2,i} \ \text{d}_1 \ast_1^{-1} U_{\perp,i} \right)  
+ U_{r,i} \wedge D_{r,1} \Theta_i \right)
- \chi_i U_{r_i} -
\xi
\left( \ast_{2,i} \ \text{d}_1 \ast_1^{-1} \text{d}_1^T \ \Theta_i 
\right.
\\
+ \frac{2}{r_i} D_{r,1} \Theta_i  
+ D_{r,2} \Theta_i \Bigl) = 0 .
\end{split}
\label{eq: wedge energy balance}
\end{equation}

\noindent where $K_i$ is a vector of $N_0$ entries containing the discrete dual 0-forms $\left(\bm{u}_\perp \cdot \bm{u}_\perp \right)^\flat$ stored on the dual nodes of $i^{\text{th}}$ spherical layer, $V_{\perp,i}$ is a vector with $N_1$ entries containing the discrete 1-forms of surface velocity (which will be described later), and $D_{r,\alpha}$ is the finite difference operator for partial derivative with respect to $r$, with $\alpha$ being the order of the derivative


\textcolor{black}{We employ the approach of Hirani \cite{hirani2003discrete} for evaluating the wedge products in preceding Eqs. (\ref{eq: wedge surface momentum balance}) $-$ (\ref{eq: wedge energy balance})}. We first simplify the second term in Eq. (\ref{eq: wedge radial momentum balance}), $U_{\perp,i} \wedge U_{r,i}$. It consists of a wedge product of a discrete dual 1-form stored on the dual edges with a discrete dual 0-form stored on the dual nodes. On each dual edge, we first evaluate the average of the dual 0-forms stored on the dual nodes forming the edge, which is given by the vector $0.5 \left( \ \left | \text{d}_1^T \right | \ U_{r,i} \ \right)$, containing dual 1-forms. \textcolor{black}{We then perform the element-wise multiplication operation} with $U_{\perp,i}$, as $A_i U_{\perp,i}$, where $A_i$ is a $N_1\times N_1$ matrix defined as $A_i = \text{diag} \left( 0.5 \ \left | \text{d}_1^T \right | \ U_{r,i} \ \right)$, $\text{diag}(.)$ denotes the diagonal matrix composed of the enclosed vector entries. Similarly, the second term in Eq. (\ref{eq: wedge energy balance}), $U_{\perp,i} \wedge \Theta_{i}$, is simplified as $C_i U_{\perp,i}$, where $C_i$ is a $N_1\times N_1$ matrix defined as $C_i = \text{diag} \left( 0.5 \ \left |\text{d}_1^T \right | \ \Theta_{i} \ \right)$.

Consider the third term in Eq. (\ref{eq: wedge radial momentum balance}), $U_{r,i} \wedge(\ast_{2,i} \ \text{d}_1 \ast_1^{-1} U_{\perp,i})$. The first operation $\ast_1^{-1} U_{\perp,i}$ maps the discrete dual 1-forms stored on the dual edges onto the primal 1-forms stored on the primal edges. \textcolor{black}{The subsequent operation} $\text{d}_1 ~ \ast_1^{-1} ~ U_{\perp,i}$ then transforms the primal 1-forms into the primal 2-forms stored on the primal triangles. Finally, $\ast_{2,i} \ \text{d}_1 \ast_1^{-1} U_{\perp,i}$ maps the primal 2-forms onto dual 0-forms stored on the dual nodes. Therefore, the wedge product, $U_{r,i} \wedge(\ast_{2,i} \ \text{d}_1 \ast_1^{-1} U_{\perp,i})$, is the element-wise multiplication of the resulting dual 0-forms with the radial velocity dual 0-forms stored in $U_{r,i}$, written as $B_i \ast_{2,i} \ \text{d}_1 \ast_1^{-1} U_{\perp,i}$, where $B_i$ is a $N_0\times N_0$ matrix defined as $B_i =  \text{diag}\left( U_{r,i} \right )$. Similarly, the third term in Eq. (\ref{eq: wedge energy balance}), $\Theta_i \wedge(\ast_{2,i} \ \text{d}_1 \ast_1^{-1} U_{\perp,i})$, is simplified as $D_i \ast_{2,i} \ \text{d}_1 \ast_1^{-1} U_{\perp,i}$, where $D_i$ is a $N_0\times N_0$ matrix defined as $D_i =  \text{diag}\left( \Theta_{i} \right )$.

Now, we simplify the convective term in Eq. (\ref{eq: wedge surface momentum balance})$,\ast_1 \left( V_{\perp,i}  \ \wedge\ \ast_{0,i}^{-1} [-\text{d}_0^{\text{T}}] \ U_{\perp,i} \right)$. Since the convective term is defined on the dual edges, $V_{\perp,i}  \ \wedge\ \ast_{0,i}^{-1} [-\text{d}_0^{\text{T}}] \ U_{\perp,i}$ is defined on the primal edges. Since $U_{\perp,i}$ consists of dual 1-forms stored on dual edges, $[-\text{d}_0^{\text{T}}] \ U_{\perp,i}$ results in dual 2-forms stored on dual cells, followed by $\ast_{0,i}^{-1} [-\text{d}_0^{\text{T}}] \ U_{\perp,i}$ consisting of primal 0-forms stored on primal nodes. Therefore, the operands of the wedge product are a 1-form and a 0-form, resulting in a 1-form defined on the primal edge. This requires the surface velocity 1-form $\bm{u}_\perp^\flat$ in the first operand to be defined on the primal edges, i.e., we denote the 1-form $v_i = \int_{\sigma_i^1} \bm{u}.\text{d}l$. The velocity 1-form $v_i$ denotes the velocity tangential to the edges of the triangles. The vector $V_{\perp,i}$ with $N_1$ entries defined earlier, therefore, consists of these velocity 1-forms defined on the primal edges. In order to obtain a linear representation of the convective term, we compute the tangential surface velocity 1-forms $v_{i}$ from previously-known dual surface velocity 1-forms through the interpolation method described in Hall et al.~\cite{hall1991dual}, which involves obtaining vector fields on the primal triangles. The convective term is finally simplified as $\ast_1 \ W_{V,i}\ \ast_{0,i}^{-1} [-\text{d}_0^{\text{T}}] U_{\perp,i}$, where $W_{V,i} = 0.5 \ \text{diag} \left( V_{\perp,i} \right)  \left | \text{d}_0 \right |$ is a sparse $N_1\times N_2$ matrix.

Also, since the orientation of the primal simplices and dual cells remains the same for all the spherical layers, the DEC operators $\text{d}_0$ and $\text{d}_1$ are not a function of the radial layers. The $n^{\text{th}}$ diagonal element of $\ast_1^{-1}$, $\frac{\ast \sigma_{n,i}^1}{\sigma_{n,i}^1}$ ($\sigma^j_{n,i}$  is the volume of primal j-simplex on $i^{\text{th}}$ spherical layer), also remains constant for all spherical layers. Therefore, we drop the subscript $i$ for $\text{d}_0$, $\text{d}_1$, and $\ast_1^{-1}$ operators. Simplifying the wedge products further, we finally obtain the generalized DEC-FD equations discretized in space and the time discretization is described later:

\begin{equation}
\begin{split}
\color{blue}  
\ast_{2,i} \ \text{d}_1  \ast_1^{-1} 
\color{black}
U_{\perp,i} 
+ 
\frac{1}{r_i^2} 
\color{red}   
D_{r,1}
\color{black}
(r_i^2 U_{r,i})  = 0 ,
\end{split}
\label{eq: DEC-FD continuity equation}
\end{equation}

\begin{equation}
\begin{split}
\frac{\partial U_{\perp,i}}{\partial t} 
+ 
\color{blue}  
\ast_1 \ W_{V,i}\ \ast_{0,i}^{-1} [-\text{d}_0^{\text{T}}] 
\color{black}
\ U_{\perp,i} 
+ 
\frac{\color{blue} A_i}{ r_i} U_{\perp,i} 
+ 
{\color{blue} A_i}  \ {\color{red} D_{r,1}} U_{\perp,i} 
+ 
{\color{blue} \text{d}_1^{\text{T}} } P_i \
+
 \frac{1}{2}  {\color{blue} \text{d}_1^{\text{T}} }K_i\
-  
\left[ \left(  
{\color{blue} \text{d}_1^T \ast_{2,i} \ \text{d}_1  \ast_1^{-1}} \right. \right.
 + \\
 \left. {\color{blue} \ast_1 \ \text{d}_0 \ast_{0,i}^{-1} [-\text{d}_0^{\text{T}}]  }
 \right)  \ U_{\perp,i}
+  \frac{2 \ U_{\perp,i}}{r_i^2} 
+ \frac{2}{r_i} \ {\color{blue} \text{d}_1^{\text{T}} } U_{r,i} 
+ \frac{2}{r_i} {\color{red} D_{r,1} }U_{\perp,i} 
+ {\color{red} D_{r,2} } U_{\perp,i}  \Bigl] = 0 ,
\end{split}
\end{equation}

\begin{equation}
\begin{split}
\frac{\partial U_{r,i}}{\partial t} 
+ 
\left ( {\color{blue} \ast_{2,i} \ \text{d}_1 \ast_1^{-1} A_i }
-
{\color{blue} B_i \ast_{2,i} \ \text{d}_1 \ast_1^{-1} } \right) U_{\perp,i} 
- 
\frac{K_i}{r_i} 
+ 
B_i \ {\color{red}  D_{r,1} } U_{r,i} 
+  
{\color{red}  D_{r,1} } P_i 
-  \zeta_i \Theta_i 
-
\left(  {\color{blue} \ast_{2,i} \ \text{d}_1 \ast_1^{-1} \text{d}_1^T } \ U_{r,i} 
-
\right.
\\
\left.
 \frac{4}{r_i} \left( {\color{blue} \ast_{2,i} \ \text{d}_1  \ast_1^{-1} }  \right)  U_{\perp,i}
- \frac{ 6 \ U_{r,i} }{r_i^2} + 
{\color{red}  D_{r,2} } U_{r,i} 
\right ) = 0  ,
\end{split}
\end{equation}

\begin{equation}
\begin{split}
\psi \left( \frac{\partial \Theta_i}{\partial t} 
+ 
\left ( {\color{blue} \ast_{2,i} \ \text{d}_1 \ast_1^{-1} C_i }  
- 
{\color{blue} D_i \ast_{2,i} \ \text{d}_1 \ast_1^{-1}} \right ) U_{\perp,i} 
+ 
B_i \  {\color{red}  D_{r,1} } \Theta_i 
\right )
- \chi_iU_{r,i}
-
\xi
 \left( {\color{blue} \ast_{2,i} \ \text{d}_1 \ast_1^{-1} \text{d}_1^T }\ \Theta_i 
 + 
 \frac{2}{r_i} {\color{red}  D_{r,1} }\Theta_i  
 + {\color{red} D_{r,2} } \Theta_i 
\right ) = 0  ,
\end{split}
\end{equation}

\noindent where the DEC operators are highlighted in blue color and FD operators are indicated in red color (refer to the web version for the colors). \textcolor{black}{In the preceding equations, the time derivative is not approximated, and FD operators are not specified. Therefore, these are the most generalized DEC-FD equations, and any time integration method and finite difference scheme can be employed.}

\begin{figure}[ht]
\centering
\begin{center}
\begin{tabular}{c}
\includegraphics[scale=0.4]{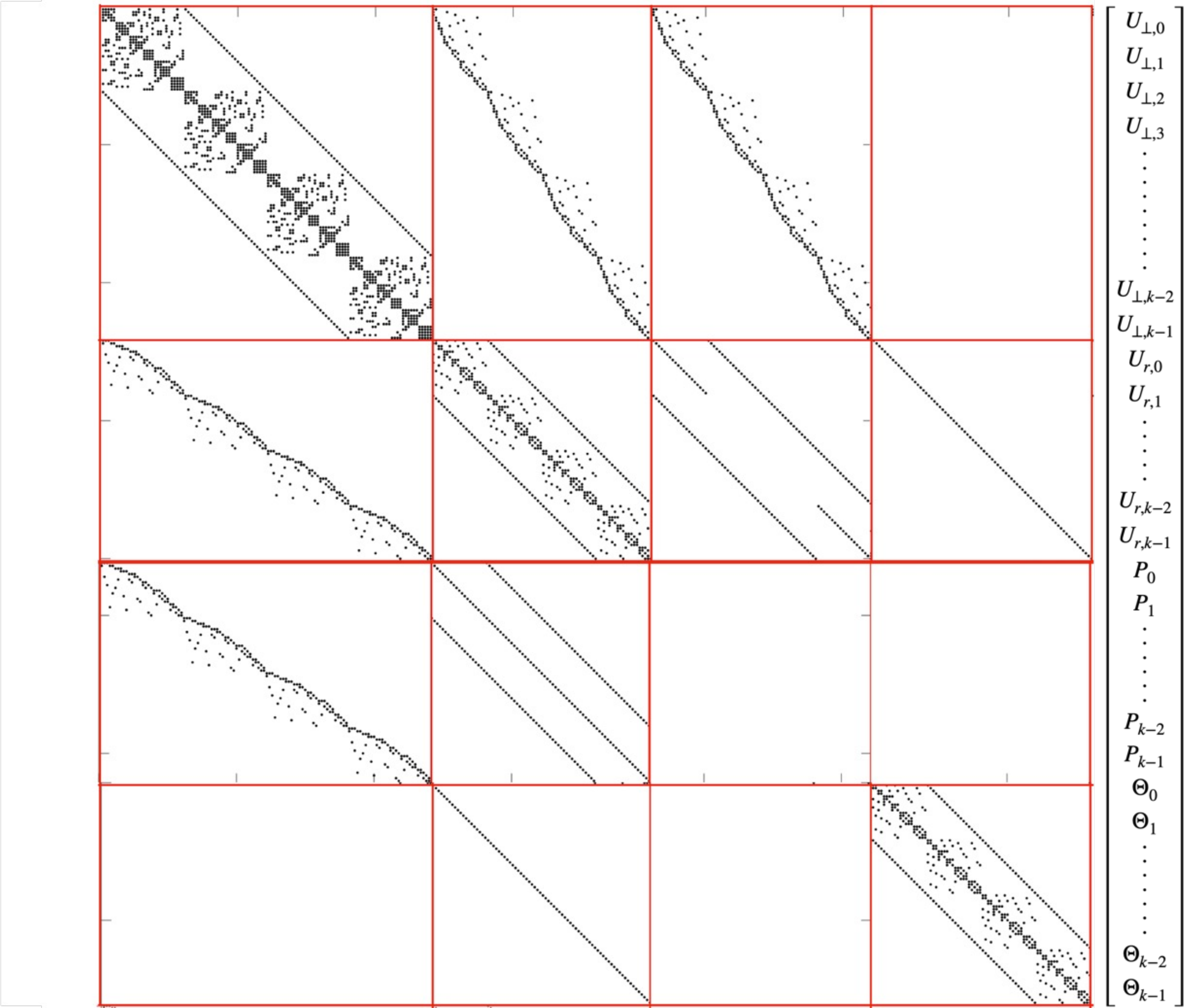} \\
\end{tabular}
\end{center}
\caption{Matrix structure and the solution vector}
\label{fig: matrix structure}
\end{figure}

In this work, we employ a second-order finite difference scheme for the radial spatial derivatives and evolve the equations in time using a semi-implicit time integration scheme, where linear terms are treated implicitly and the non-linear terms are treated explicitly. The discretized equations are further simplified as:

\begin{equation}
\begin{split}
\ast_{2,i} \ \text{d}_1  \ast_1^{-1} U_{\perp,i}^{n+1} + \frac{1}{2 \Delta r} U_{r,i+1}^{n+1} - \frac{1}{2 \Delta r} U_{r,i-1}^{n+1} +  \frac{2}{r_i} U_{r,i}^{n+1} = 0  ,
\end{split}
\label{eq: final decfd continuity}
\end{equation}

\begin{equation}
\begin{split}
\left[
\left(
\frac{1}{\Delta t}
- \frac{2}{r_i^2}
+ \frac{2}{(\Delta r)^2}
\right)
\mathbb{I} 
- \text{d}_1^T \ast_{2,i} \ \text{d}_1  \ast_1^{-1}
- \ast_1 \ \text{d}_0 \ast_{0,i}^{-1} [-\text{d}_0^{\text{T}}] 
\right]
U_{\perp,i}^{n+1} 
- \left[ \frac{1}{\Delta r} \left( \frac{1}{r_i} +  \frac{1}{\Delta r} \right) \mathbb{I}  \right]U_{\perp,i+1}^{n+1}  
+ \\ 
\left[ \frac{1}{\Delta r}\left( \frac{1}{r_i} -  \frac{1}{\Delta r} \right) \mathbb{I}  \right] U_{\perp,i-1}^{n+1}
- \frac{2}{r_i}  \text{d}_1^T U_{r,i}^{n+1} 
+ \text{d}_1^T P_i^{n+1} 
= 
 \left( \frac{1}{\Delta t} \mathbb{I} 
 - \ast_1 \ W_V^n\ \ast_{0,i}^{-1} [-\text{d}_0^{\text{T}}] 
 - \frac{1}{r_i} A_i^n \right) \ U_{\perp,i}^n  - \\
 \frac{1}{2\Delta r} A_i^n U_{\perp,i+1}^n 
 + \frac{1}{2\Delta r} A_i^n U_{\perp,i-1}^n 
 - \frac{1}{2} \text{d}_1^T K_i^n  ,
\end{split}
\end{equation}


\begin{equation}
\begin{split}
\left[
\left( 
\frac{1}{\Delta t}
+ \frac{ 6} {r_i^2}
+ \frac{2}{(\Delta r)^2} 
\right) 
\mathbb{J}
- \ast_{2,i} \ \text{d}_1 \ast_1^{-1} \text{d}_1^T 
\right]
U_{r,i}^ {n+1} 
+ \frac{4}{r_i} \ast_{2,i}  \text{d}_1 \ast_1^{-1} U_{\perp,i}^{n+1} 
+ \frac{1}{2\Delta r} P_{i+1}^{n+1} 
- \frac{1}{2\Delta r} P_{i-1}^{n+1} 
- 
\\
\zeta_i \Theta_i^{n+1}  
- \frac{1}{(\Delta r)^2} U_{r,i+1}^{n+1} 
- \frac{1}{(\Delta r)^2} U_{r,i-1}^{n+1} 
= \frac{1}{\Delta t} U_{r,i}^n 
+ 
\left( 
- \ast_{2,i} \ \text{d}_1 \ast_1^{-1} A_i^n \ + B_i^n \ast_{2,i} \ \text{d}_1 \ast_1^{-1}  
\right ) 
U_{\perp,i}^n 
+ 
\\
\frac{1}{r_i}  K_i^n
- \frac{1}{2 \ \Delta r} B_i^n \ U_{r,i+1}^n 
+ \frac{1}{2 \ \Delta r} B_i^n \ U_{r,i-1}^n  ,
\end{split}
\end{equation}

\begin{equation}
\begin{split}
\left[
\left(
\frac{\psi}{\Delta t}
+ \frac{2 \xi}{(\Delta r)^2}
\right)
\mathbb{J} 
- \xi \ast_{2,i} \ \text{d}_1 \ast_1^{-1} \text{d}_1^T
\right]
\Theta_i^{n+1} 
- \frac{\xi}{\Delta r} \left(\frac{1}{r_i} + \frac{1}{\Delta r} \right) \Theta_{i+1}^{n+1} 
+ \frac{\xi}{\Delta r} \left(\frac{1}{r_i} - \frac{1}{\Delta r} \right) \Theta_{i-1}^{n+1} 
- \chi_i U_{r,i}^{n+1}
= \\
\frac{\psi}{\Delta t} \Theta_i^n 
+ 
\psi \left( 
- \ast_{2,i} \ \text{d}_1 \ast_1^{-1} \ C_i^n  + D_i^n \ast_{2,i} \ \text{d}_1 \ast_1^{-1} 
\right ) 
U_{\perp,i}^n 
- \frac{\psi}{2 \ \Delta r} B_i^n \ \Theta_{i+1}^n 
+ \frac{\psi}{2 \ \Delta r} B_i^n \ \Theta_{i-1}^n   ,
\end{split}
\label{eq: final decfd energy}
\end{equation}

\noindent where the superscript $n$ denotes the current time level, $\Delta r$ is the spacing of the grid points in the radial direction, $\Delta t$ is the time step, and $\mathbb{I}$ and $\mathbb{J}$ are $N_1 \times N_1$ and  $N_0 \times N_0$  identity matrices, respectively. Preceding Eqs. (\ref{eq: final decfd continuity}) $-$ (\ref{eq: final decfd energy}) are solved as a fully coupled system $Mx = b$, where the structure of $(N_1 + 3N_0)k \times (N_1 + 3N_0)k$ sparse matrix $M$ and the corresponding solution vector $x$ consisting of subvectors 
\begin{equation*}
U_{\perp,0}, U_{\perp,1} \ldots, U_{\perp,k-1}, U_{r,0}, U_{r,1} \ldots, U_{r,k-1}, P_{0}, P_{1} \ldots, P_{k-1},\Theta_{0}, \Theta_{1} \ldots, \Theta_{k-1}
\end{equation*}
are shown in the Fig. \ref{fig: matrix structure}, and $b$ is a vector of $(N_1 + 3N_0)k$ entries.


\section{Numerical experiments}
\label{section: Numerical Experiments}

The DEC-FD MPI code used in this work is developed in the Fluid and Plasma Simulation Laboratory (FPSL) at King Abdullah University of Science and Technology (KAUST), based on the Portable and Extensible Toolkit for Scientific computing (PETSc) framework \cite{petsc-user-ref,petsc-efficient,KarypisKumar98,kirby2004}. The base surface mesh is handled using PETSc's mesh topology abstraction DMPLEX \cite{knepley2009mesh}. We employ MUltifrontal Massively Parallel sparse direct Solver (MUMPS)\cite{MUMPS01,MUMPS02} interfaced with PETSc for the linear solve, with relative tolerance being $10^{-11}$. In the following, we corroborate the proposed DEC-FD methodology with numerical test cases:

\begin{enumerate}
  \item Quantification of the critical Rayleigh number for spherical shells with $0.2 \leq \eta \leq 0.8$ subject to internal heating, where $\eta=R_0/R_1$.
  \item \textcolor{black}{Simulation of robust convective patterns including the} formation of a giant spiral roll, covering all the spherical surface, in a moderately thin spherical shell ($\eta = 0.8$) subject to internal heating.
  \item Quantification of the output parameters for RBC in spherical shell with $\eta = 0.6$, $1.5\times10^3 \leq Ra \leq 3\times10^4$.
\end{enumerate}

For all the test cases, the boundaries are rigid, non-slip, and at constant temperatures, which yield $U_\perp = 0, U_r = 0,$ and $\Theta = 0$ at $r = R_0$ and $r = R_1$. Table \ref{table: simulation details} summarizes the details of the simulation runs consisting of 2 categories. A majority of the calculations in this work belong to category 1, which involves solving for 5.5 million DOF (Degrees of freedom) in thick and moderately thick spherical shells ($0.2 \leq \eta \leq 0.6$). In order to resolve the high wave number features in moderately thin spherical shells ($\eta=0.8$) and demonstrate the code's capability of handling higher grid resolutions, we performed calculations solving for 14.7 million DOF that belong to category 2. We note that the memory requirement of the direct solver is the bottleneck for simulation runs with higher grid resolutions. \textcolor{black}{Also}, since the aim of this work is to present the DEC-FD hybrid methodology \textcolor{black}{and verify it}, parallelization and solver are of no prime importance in this paper. 


\begin{table}[ht!]
\caption{Simulation details}
\vspace{0.2cm}
\begin{center}
\begin{tabular}{l l l}
\hline
\hline
& \\
& Category 1 & Category 2 \\
& \\
  
Number of compute nodes & 8 & 16\\
 
Surface mesh refinement level $(s_r)$ & 5 & 6 \\

Number of spherical layers excluding boundary spherical layers $(k)$ & 60 & 40 \\

\textcolor{black}{Number of elements on each spherical layer }$\left( N_0 \right)$ & 20 480 & 81 920\\

Degrees of freedom (DOF) & 5 529 600 & 14 745 600\\

Aspect Ratio ($\eta=R_0/R_1$) & 0.2, 0.4, 0.5, 0.6 & 0.8 \\

& \\
\hline
\end{tabular}
\end{center}
\label{table: simulation details}
\end{table}
\FloatBarrier
\textcolor{black}{The aforementioned test cases can be grouped into two classes based on how the radial temperature gradient is imposed: (\romannum{1}) internal heating, (\romannum{2}) basal heating. The aforementioned test cases 1 and 2 fall in class (\romannum{1}) and test case 3 belongs to class (\romannum{2}).}

\subsection{Internal heating}

For all the test cases in this subsection, we adopt the convection model proposed by Chandrasekhar \cite{chandrasekhar2013hydrodynamic}. The \textcolor{black}{fluid in} spherical shell has uniform distribution of heat sources and is subject to a gravitational field $\bm{g} = - \gamma r \  \bm{e_r}$ and radial conductive temperature gradient, $\nabla T_c(r) = -\beta r \  \bm{e_r}$,  where $g(r) = - \gamma r$, $f(r) = -\beta r$, and $\gamma$ and $\beta$ are positive constants. We employ the outer radius $R_1$ as the length scale, $R_1^2/ \nu$ as the time scale, $\beta R_1^2 \nu/ \kappa$ as the unit of temperature fluctuation, which result in the non-dimensional control parameters Rayleigh number $C_l$, the Prandtl number $Pr$, and the aspect ratio $\eta$, defined by

\begin{equation}
C_l = \frac{\alpha \beta \gamma }{\kappa \nu} R_1^6, ~~~ Pr = \frac{\nu}{\kappa}, ~~~  \eta = \frac{R_0}{R_1}.
\label{eq: non-dimensional control paramters}
\end{equation}

\noindent We note that the notation for Rayleigh number is $C_l$ in this subsection, following the convention of \cite{chandrasekhar2013hydrodynamic}. Based on this non-dimensionalization and preceding definitions for $g(r)$ and $f(r)$, the coefficients in Eqs. (\ref{eq: non-dimensional continuity}) $-$ (\ref{eq: non-dimensional energy equation}) turn out to be 

\begin{equation}
\zeta(r) = C_l r, ~~~ \psi = Pr, ~~~ \chi(r) = r, ~~~  \xi = 1.
\end{equation}

\subsubsection{Quantification of critical Rayleigh number}
\label{subsection: Quantification of critical Rayleigh number}

\begin{figure}[ht]
\centering
\begin{center}
\begin{tabular}{c}
\includegraphics[scale=1.2]{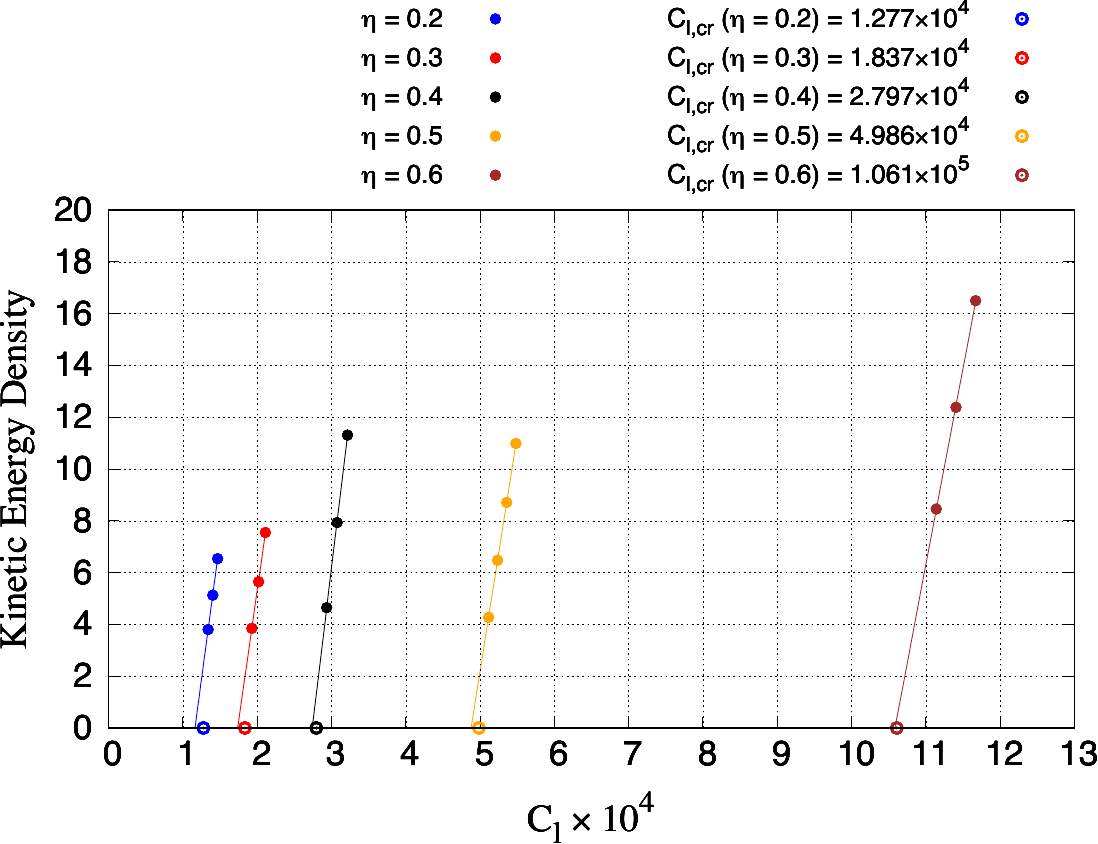} \\
\end{tabular}
\end{center}
\caption{Fluid kinetic energy density at different stationary states near the onset of convection. The stationary fluid is perturbed with spherical harmonic modes $\bm{l = 0-10}$. The filled circles are numerically obtained from the present study and the unfilled circles are the analyitcal critical Rayleigh numbers derived by Chandrasekhar \cite{chandrasekhar2013hydrodynamic}.}
\label{fig: numerical critical rayleigh number}
\end{figure}
\begin{figure}[ht!]
\centering
\begin{center}
\begin{tabular}{c}
\includegraphics[scale=1.0]{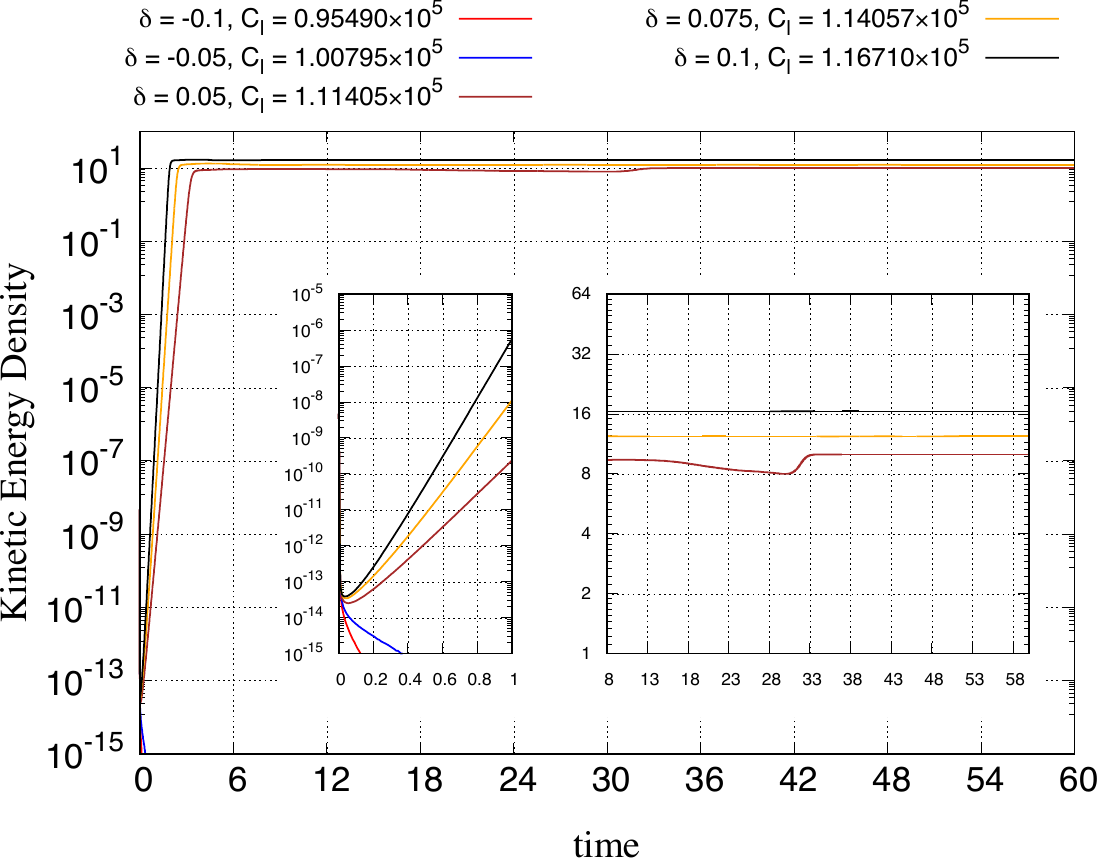} \\
\end{tabular}
\end{center}
\caption{Evolution of fluid kinetic energy density for $\bm{\delta = \frac{C_{l} - C_{l,cr}}{C_{l,cr}} =  -0.1, -0.05, 0.05, 0.075,}$ and $\bm{0.1}$, where $\bm{C_l}$ is the Rayleigh number and $\bm{C_{l,cr}}$ is the analytical critical Rayleigh number. The stationary fluid is perturbed with spherical harmonic modes $\bm{l = 0-10}$ for $\bm{\eta = 0.6}$.}
\label{fig: 0.6eta_ke_evolution}
\end{figure}

When small disturbances are introduced \textcolor{black}{in a static fluid with pure heat conduction in a spherical shell}, the instability must set in, when $C_l$ exceeds a certain critical value $C_{l,cr}$, resulting in the onset of convection. When $C_l < C_{l,cr}$, 
the perturbed kinetic energy decays exponentially to return to the static state. When $C_l$ just exceeds $C_{l,cr}$, the critical spherical harmonic mode $l_c$ triggers the instability, and the fluid's kinetic energy grows exponentially to reach stationary states. One of the primary goals of this work is to estimate $C_{l,cr}$. Numerically, in order to onset convection, infinitesimally small probabilistic perturbations are introduced in radial velocity,

\begin{equation}
\begin{split}
u_r = & \sum_{\ell = 0}^{l_{\text{max}}} \sum_{m = -\ell}^{m = \ell} 2 \ \epsilon  \ \Re \left \{\alpha_{\ell m}  \ Y_{\ell m}  \  \sin \left[\frac{\pi \ \left(r - r_i \right)}{ \left(r_o-r_i \right)} \right] \right \}, \\
\alpha_{\ell m} = & \frac {\mathcal{N} \left( 0,1 \right) + i \mathcal{N} \left (0,1 \right)}{\sqrt{2}},
\end{split}
\label{eq: perturbation}
\end{equation}

\noindent where $\mathcal{N}(0,1)$ is a Gaussian distribution with zero mean and unit variance. In Fig. \ref{fig: numerical critical rayleigh number}, we present the steady-state kinetic energy density,  $\left( KE =  \frac{1}{V}\int \frac{u^2}{2} \text{d} V \right .$, where $V$ is the volume of the spherical shell$\Bigl )$, attained for $C_l$ just exceeding $C_{l,cr}$ for spherical shells with $\eta = 0.2, 0.3, 0.4, 0.5,$ and $0.6$. For $C_l$ close to $C_{l,cr}$, the kinetic energy density varies linearly with the Rayleigh number \cite{al2004varying}. Using this fact, the numerically-estimated critical Rayleigh number, $\hat{C}_{l,cr}$, is obtained by extrapolating the steady-state kinetic energy density  attained at a few Rayleigh numbers to zero kinetic energy density. To substantiate that the extrapolated values are indeed critical, we plot the evolution of kinetic energy density for the case of $\eta = 0.6$, shown in the Fig. \ref{fig: 0.6eta_ke_evolution}. We observe that for $C_l < \hat{C}_{l,cr} = 1.0585\times 10^5$, the kinetic energy decays to zero, while for $C_l > \hat{C}_{l,cr}$, the kinetic energy density \textcolor{black}{grows and} attains a non-zero value. 

\begin{figure}[ht!]
\centering
\begin{center}
\begin{tabular}{cc}
\includegraphics[scale=0.18]{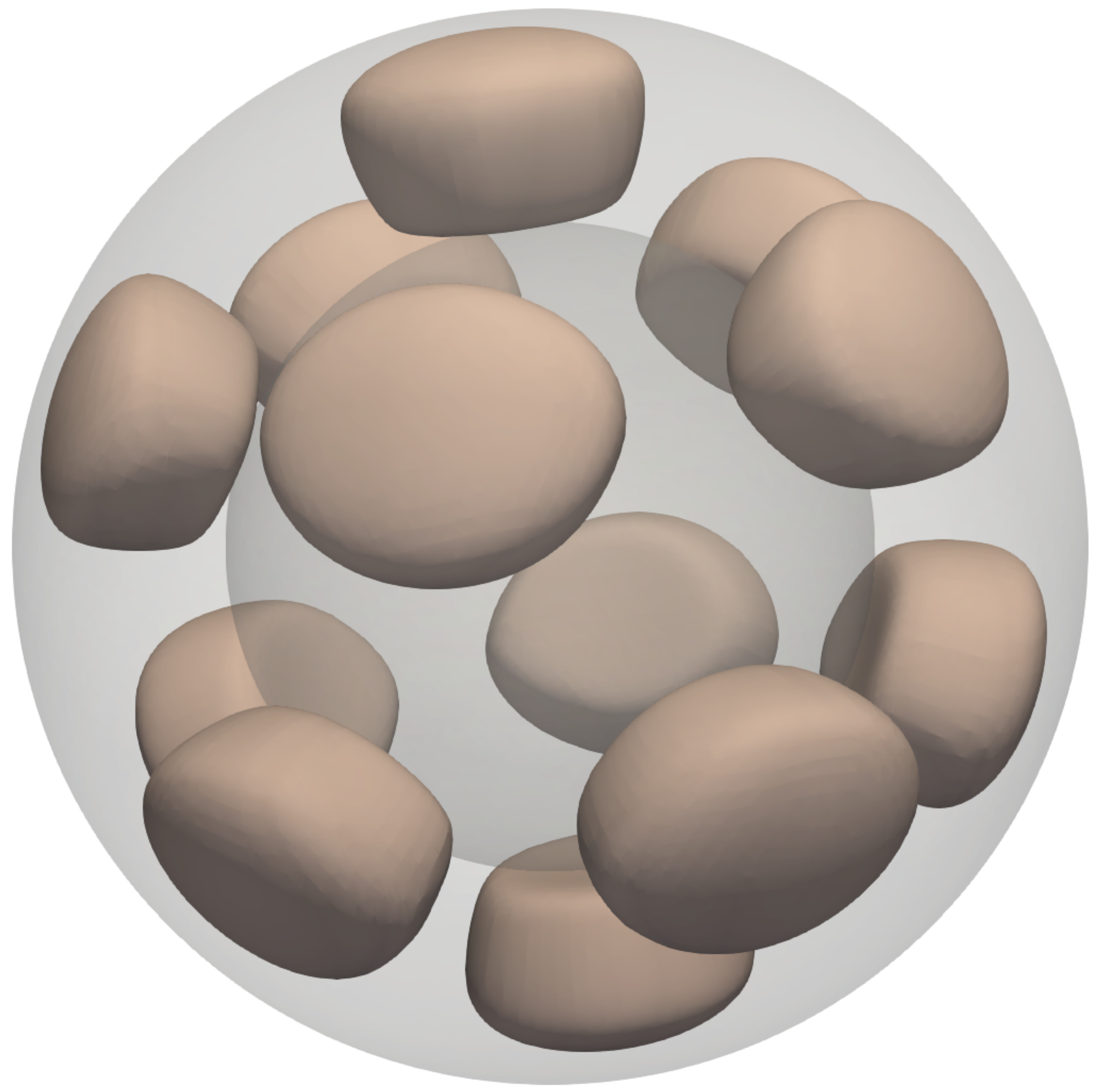} &
\includegraphics[scale=0.32]{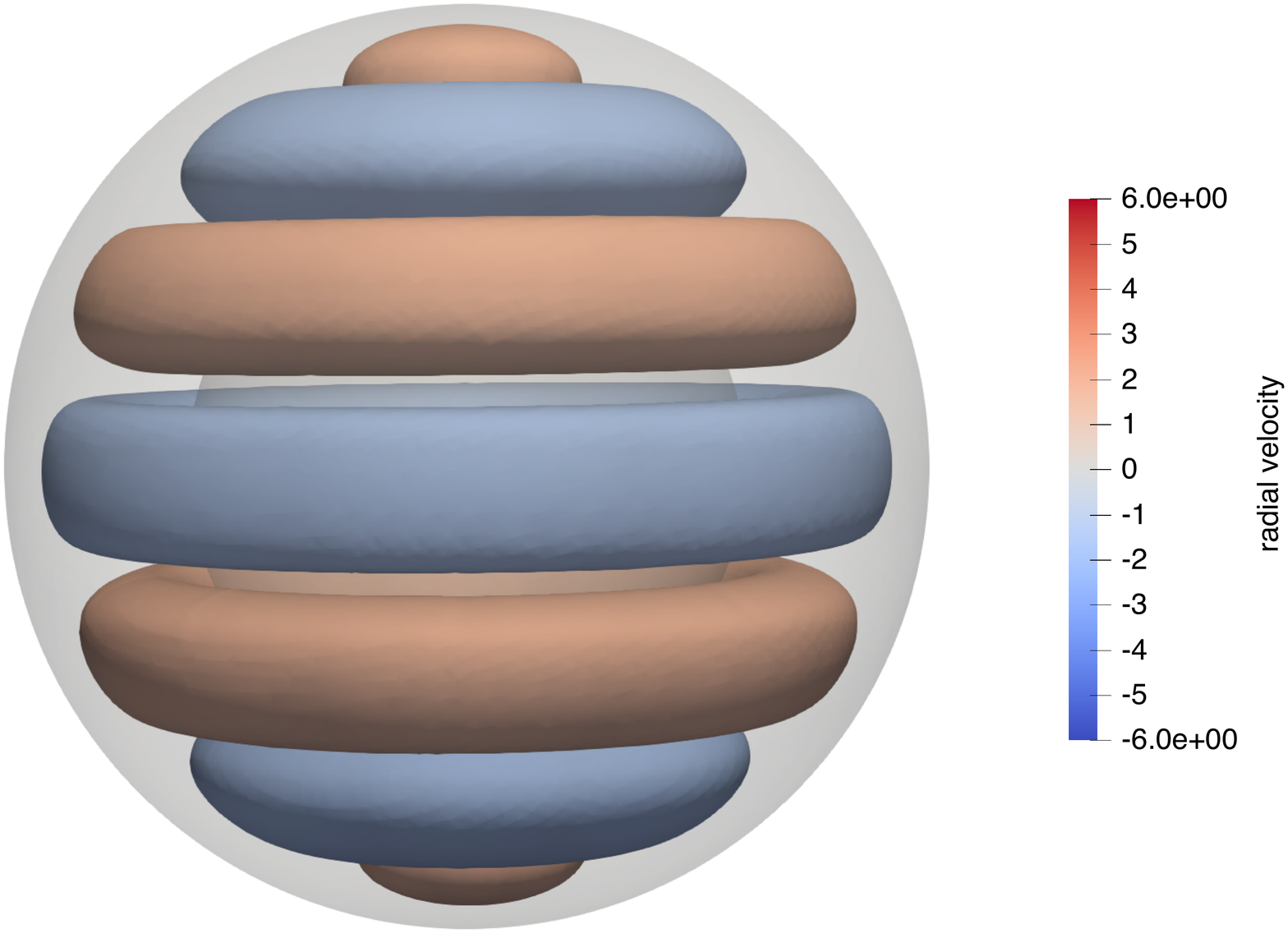}\\
(a) & (b)\\
\end{tabular}
\end{center}
\caption{(a) Dodecahedron convection pattern for $\bm{\eta = 0.6}$ visualized by the contours of the positive radial velocity; (b) stable axisymemtric convective pattern for $\bm{\eta = 0.6}$ visualized by the contours of the radial velocity.}
\label{fig: dodecahedron pattern}
\end{figure}

\textcolor{black}{We take a small digression here. In Fig. \ref{fig: 0.6eta_ke_evolution}, we observe a kink in the KE evolution for $C_l = 1.11405\times 10^5$. In order to explain this behavior, we plot the radial velocity contours before and after the kink. The perturbations to the static state, which are free from any symmetry, first result in a steady-state dodecahedral symmetric pattern as shown in Fig. \ref{fig: dodecahedron pattern}(a) with upwelling plumes. For spherical harmonic mode $l=6$, Busse \cite{busse_1975} predicted the formation of a symmetric dodecahedral pattern. Also, from linear theory, Chandrasekhar \cite{chandrasekhar2013hydrodynamic} analytically predicted $l=6$ being the unstable mode for $\eta = 0.6$. Therefore, it can be argued that $l=6$ is the unstable mode that is triggered in the simulation, resulting in the dodecahedron pattern. However, this pattern is unstable and it is broken at $t=31$ to form an axisymmetric pattern as shown in Fig. \ref{fig: dodecahedron pattern}(b). This breaking of the unsteady dodecahedron pattern to a more stable branch is in line with Arrial et al. \cite{arrial2014sensitivity}.}

\begin{figure}[ht]
\centering
\begin{center}
\begin{tabular}{c}
\includegraphics[scale=1.0]{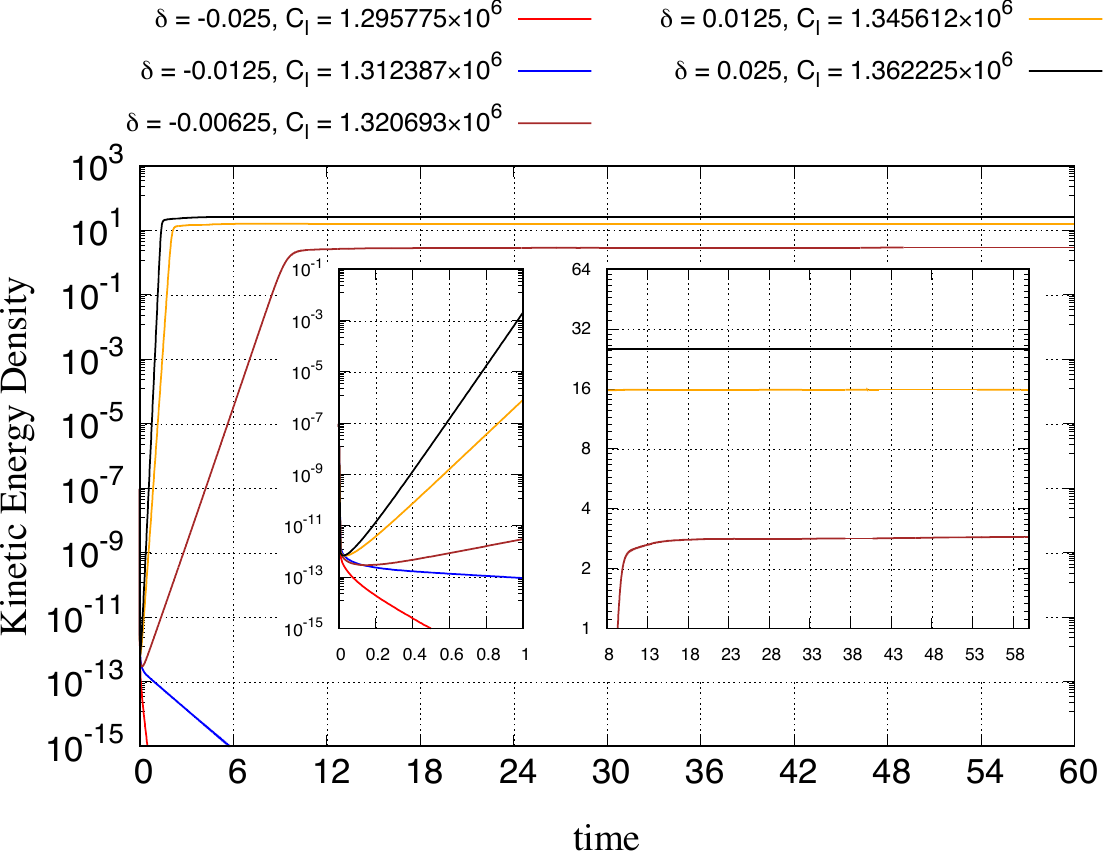} \\
\end{tabular}
\end{center}
\caption{Evolution of fluid kinetic energy density for $\bm{\delta = \frac{C_{l} - C_{l,cr}}{C_{l,cr}} =  -0.025, -0.0125, -0.00625, 0.0125,}$ and $\bm{0.025}$, where $\bm{C_l}$ is the Rayleigh number and $\bm{C_{l,cr}}$ is the analytical critical Rayleigh number. The stationary fluid is perturbed with spherical harmonic modes $\bm{l = 0-20}$ for $\bm{\eta = 0.8}$.}
\label{fig: 0.8eta_ke_evolution}
\end{figure}

In the case of a moderately thin spherical shell, for $\eta = 0.8$, although KE increases monotonically with $C_l$, a linear trend is not observed due to the coupling of the spherical harmonic modes resulting in a number of convective patterns; see Li et al. \cite{li2005multiplicity}. Therefore, we predict $\hat{C}_{l,cr}$ through the temporal evolution of $KE$, as shown in the Fig. \ref{fig: 0.8eta_ke_evolution}. From the plot, it is apparent that $1.312387\times 10^6 \leq \hat{C}_{l,cr}\leq 1.320693\times10^6$; the lower limit is considered as $\hat{C}_{l,cr}$. 

\textcolor{black}{Table \ref{table: comparison critical rayleigh numbers} shows the comparison of the numerically estimated critical Rayleigh numbers against the analytical values derived by Chandrasekhar \cite{chandrasekhar2013hydrodynamic} for spherical shells with rigid walls \textcolor{black}{for all aspect ratios}.} The numerical results are in very good agreement with the linear theory with a maximum error of less than 9\%. For $0.2 \leq \eta \leq 0.6$, the simulation runs belong to category 1, and for $\eta = 0.8$, simulation runs belong to category 2; \textcolor{black}{recall the two categories of simulations mentioned in the Table \ref{table: simulation details}}. On account of fixed resolution in the radial direction, with increasing $\eta$ from 0.2 to 0.6, the spherical shell thickness reduces, resulting in an increasing radial resolution. Because of this, we observe a decrease in the errors, maximum being $8.83\%$ for $\eta = 0.2$ and the minimum being $0.23\%$ for $\eta = 0.6$.

\begin{figure}[htp!]
\centering
\begin{center}
\begin{tabular}{c}
\includegraphics[scale=1.0]{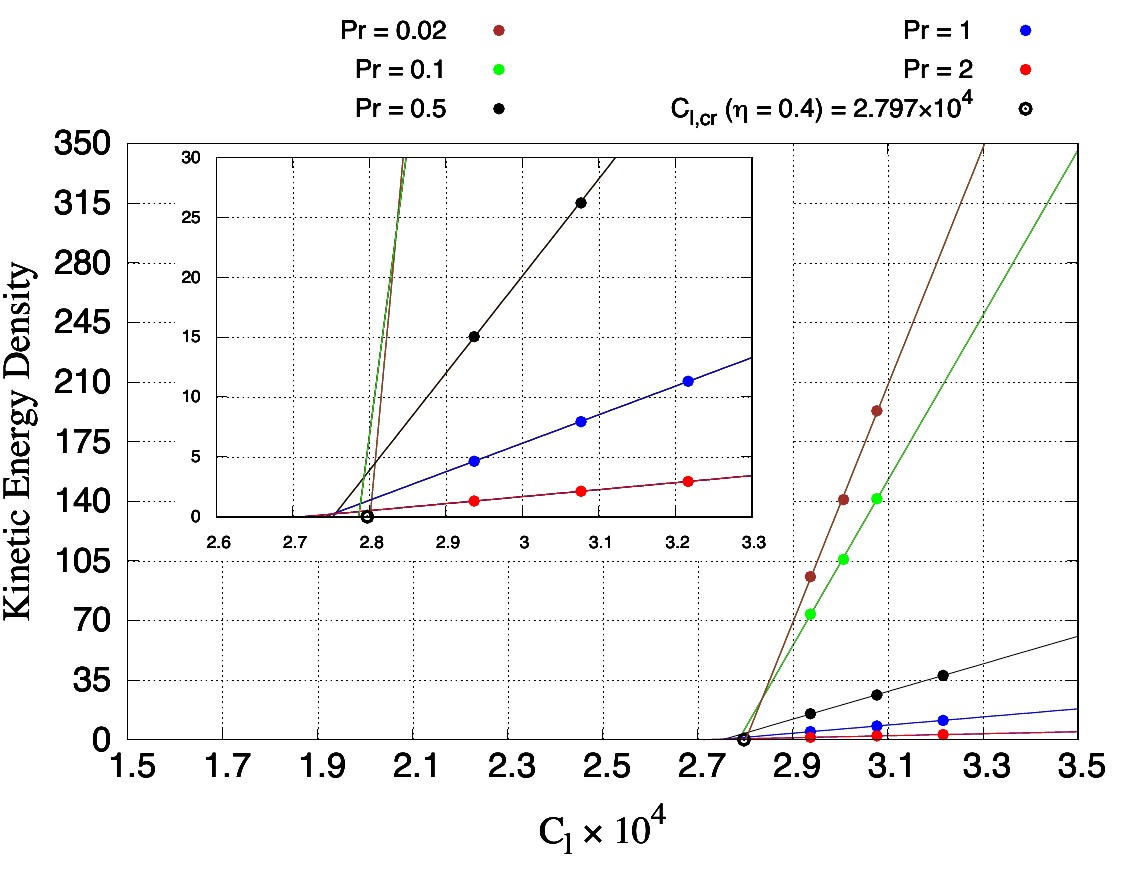} \\
\end{tabular}
\end{center}
\caption{ The critical Rayleigh number $\bm{\hat{C}_{l,cr}}$ for Prandtl numbers Pr = 0.02, 0.1, 0.5, 1, and 2, obtained by extrapolating the fluid kinetic energy density at different stationary states (shown by the filled circles) near the onset of convection. The static fluid is perturbed with spherical harmonic modes $\bm{l = 0-10}$ for spherical shell with aspect ratio $\bm{\eta = 0.4}$. The analytical critical Rayleigh number derived by Chandrasekhar \cite{chandrasekhar2013hydrodynamic} is shown by the unfilled circle.}
\label{fig: prandtl independence}
\end{figure}

Furthermore, in the normal mode analysis from linear theory, for the case when both $\beta$ and product $\alpha\gamma$ are constants, Chandrasekhar \cite[pp. 227]{chandrasekhar2013hydrodynamic} proved the principle of the exchange of stabilities is valid, where the onset of the instability must occur via a marginal state which is a stationary state. Enforcing this condition annihilates the dependence of $C_{l,cr}$ on the Prandtl number. We verify this numerically by comparing the $\hat{C}_{l,cr}$ obtained with $Pr = 0.02, 0.1, 0.5, 1,$ and $2,$ for spherical shell with $\eta = 0.4$. Figure \ref{fig: prandtl independence} shows the extrapolation of the KE at different stationary states to zero kinetic energy density to estimate the $\hat{C}_{l,cr}$. The $\hat{C}_{l,cr}$ obtained for the simulated Prandtl numbers are quantified in Table \ref{table: prandtl independence}, and a maximum error of $2.97\%$ in comparison with \cite{chandrasekhar2013hydrodynamic} is observed.

\FloatBarrier

\FloatBarrier

\FloatBarrier


\FloatBarrier

\FloatBarrier

\begin{table}[htp!]
\caption{The critical Rayleigh number $\bm{\hat{C}_{l,cr}}$ predicted numerically for different aspect ratios $\bm{\eta}= R_0/R_1$. The percentages in relative error in comparison to the analytical critical Rayleigh number $\bm{C_{l,cr}}$ derived by Chandrasekhar \cite{chandrasekhar2013hydrodynamic} are mentioned in the parentheses.}
\begin{center}
\setlength{\tabcolsep}{15pt}
\renewcommand{\arraystretch}{1.5}
\begin{tabular}{c c c c}
\hline
\hline
$\eta$ &  Simulation Category &  $\hat{C}_{l,cr}$ & $C_{l,cr}$  \\

 & & Present study & Chandrasekhar \cite{chandrasekhar2013hydrodynamic}\\
\hline 
0.2 & 1 & 1.1642$\times 10^4$ (8.83\%) & 1.277 $\times 10^4$ \\

0.3 & 1 & 1.7388$\times 10^4$ (5.34\%) & 1.837 $\times 10^4$ \\

0.4 & 1 & 2.7425$\times 10^4$ (1.94\%) & 2.797 $\times 10^4$ \\

0.5 & 1 & 4.8807$\times 10^4$ (2.11\%) & 4.986 $\times 10^4$ \\

0.6 & 1 & 1.0585$\times 10^5$ (0.23\%) & 1.061 $\times 10^5$ \\

0.8 & 2 & 1.3123$\times 10^6$ (1.27\%) & 1.329 $\times 10^6$ \\

\hline
\end{tabular}
\end{center}
\label{table: comparison critical rayleigh numbers}
\end{table}
\FloatBarrier


\begin{table}[ht!]
\caption{The critical Rayleigh number $\bm{\hat{C}_{l,cr}}$ predicted numerically for Prandtl numbers Pr = 0.02, 0.1, 0.5, 1, and 2. We report the percentages in relative error in comparison to the analytical $\bm{C_{l,cr} =  2.797 \times 10^4}$ derived by Chandrasekhar \cite{chandrasekhar2013hydrodynamic}.}
\begin{center}
\setlength{\tabcolsep}{15pt}
\renewcommand{\arraystretch}{1.5}
\begin{tabular}{c c c c c c}
\hline
\hline

\hline 
$Pr$ & 0.02 & 0.1 & 0.5 & 1 & 2 \\

$\hat{C}_{l,cr}$ $/10^4$ & 2.8010 & 2.7960 &2.7523& 2.7425 & 2.7139 \\

Error & $0.14\%$ & $ 0.03\%$ & $1.59\%$ & $ 1.94\%$ & $2.97\%$ \\

\hline
\end{tabular}
\end{center}
\label{table: prandtl independence}
\end{table}
\FloatBarrier


\subsubsection{Convergence}
We demonstrate the convergence of the DEC-FD method by performing two tests. In the first test, we simulate convection in a spherical shell with $\eta = 0.8$ for $C_l = 1.3954\times10^6$ and $Pr = 1$. By considering the solution with grid resolution $s_r = 6$ and $k= 32$, \textcolor{black}{(the highest grid resolution in our simulations)}, as the true solution, we vary the surface resolution $s_r$ from 2 to 4 and hold the number of surface layers in the radial direction constant. For this test case, deterministic perturbations are introduced in the static state to obtain similar stationary states for all the mesh resolutions. Table \ref{table: convergence test 1} shows the error in the kinetic energy density as a function of the maximum primal edge length $h$. \textcolor{black}{With the primal edge length \textcolor{black}{being reduced} by a factor of 2, we observe the error reducing by approximately a factor of 6, resulting in slightly more than second-order convergence. We observe this behaviour because the stationary structures are already well resolved near the onset of convection. Performing the convergence tests in a fully non-linear regime may slightly reduce the order of convergence.}

\begin{figure}[th!]
\centering
\begin{center}
\begin{tabular}{c}
\includegraphics[scale=1.0]{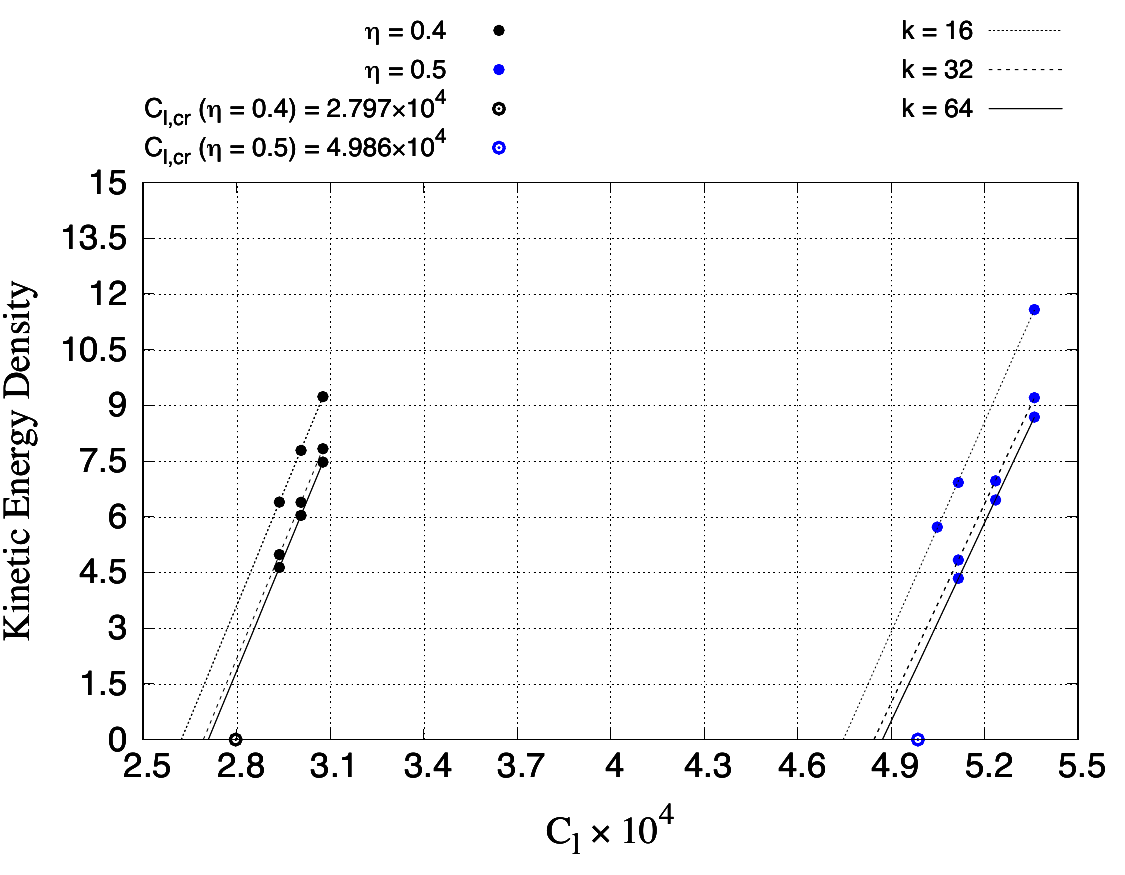} \\
\end{tabular}
\end{center}
\caption{ The critical Rayleigh number $\bm{\hat{C}_{l,cr}}$ obtained by extrapolating the fluid kinetic energy density at different stationary states near the onset of convection for radial grid resolutions $\bm{k=16, 32,}$ and 64. The stationary fluid is perturbed with spherical harmonic modes $\bm{l = 0-10}$ for aspect ratio $\bm{\eta = 0.4}$ and $\bm{0.5}$. The analytical $\bm{C_{l,cr}}$ is shown by the unfilled circle. \textcolor{black}{The critical Rayleigh number is better estimated with increasing radial mesh resolution.}}
\label{fig: convergence test 2}
\end{figure}

In the second test, we vary the number of spherical layers in the radial direction while keeping the surface resolution $s_r=5$ constant. Figure~\ref{fig: convergence test 2} shows the kinetic energy density as a function of the Rayleigh number for spherical shells with $\eta = 0.4$ and $0.5$. We extrapolate the steady-state KE to zero kinetic energy density to obtain the critical Rayleigh number as described in Section \ref{subsection: Quantification of critical Rayleigh number}. We distinguish the radial mesh resolutions $k = 16, 32,$ and 64 with different line types. We observe that with increasing grid resolution in the radial direction, the numerically-predicted critical Rayleigh number is better estimated.

\FloatBarrier
\begin{table}[htp!]
\caption{Error in steady state kinetic energy density versus the maximum surface primal edge length h }
\begin{center}
\setlength{\tabcolsep}{25pt}
\renewcommand{\arraystretch}{1.2}
\begin{tabular}{c c c c }
\hline
\hline 
h & 0.4037 & 0.2045 & 0.1026 \\
Error & 20.8264 & 3.2813 & 0.5045 \\
\hline
\end{tabular}
\end{center}
\label{table: convergence test 1}
\end{table}
 \FloatBarrier

\FloatBarrier

 \FloatBarrier
 
\subsubsection{Convective patterns}
\label{subsection: Convective patterns}
Although a plethora of convective patterns are observed in our numerical experiments for the wide range of aspect ratios that we explored, for the sake of qualitative verification, we report a few steady-state convective patterns which are in line with the literature. 

\begin{figure}[ht]
\begin{longtable}{c c}
\includegraphics[scale=0.15]{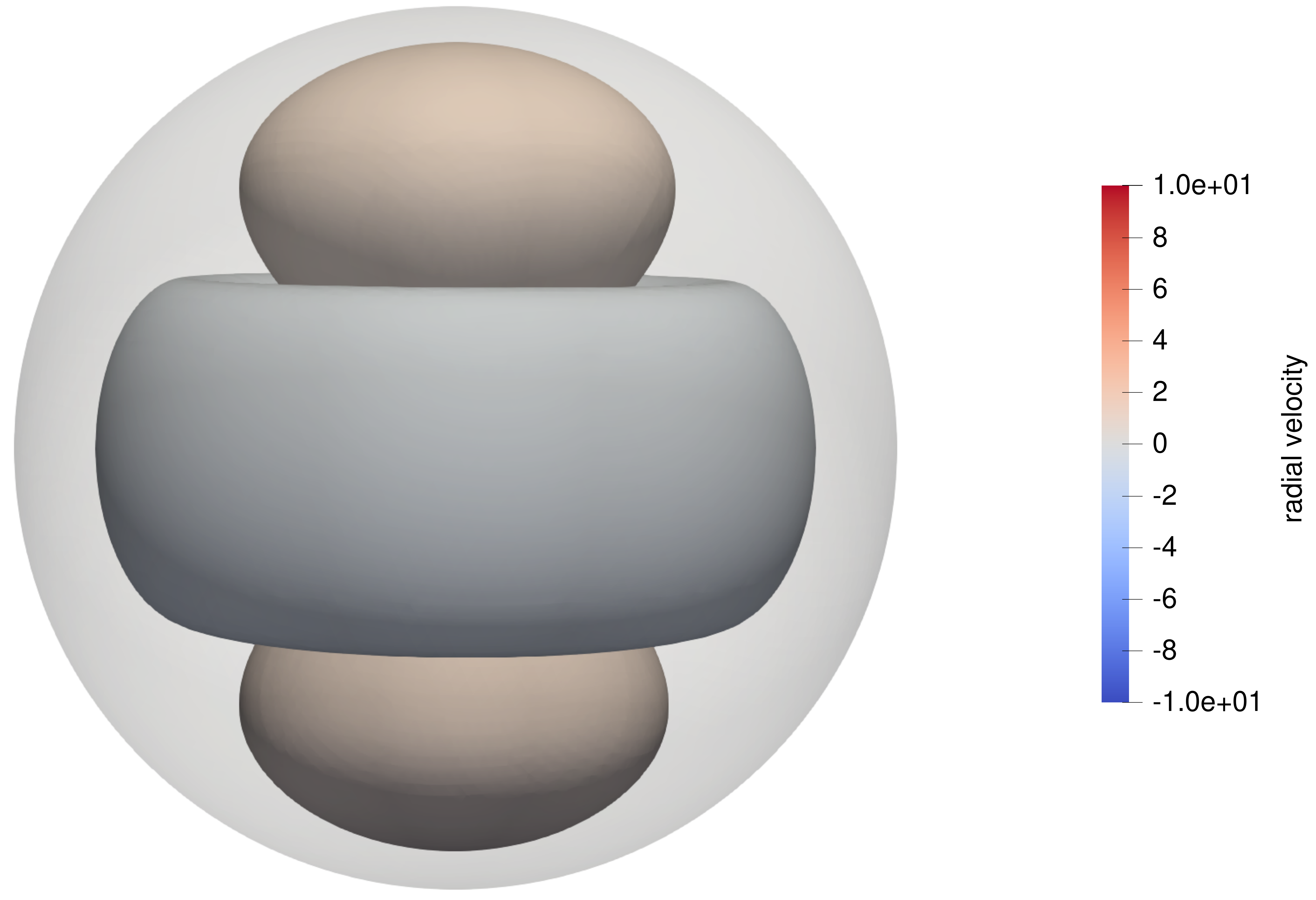} &
\includegraphics[scale=0.12]{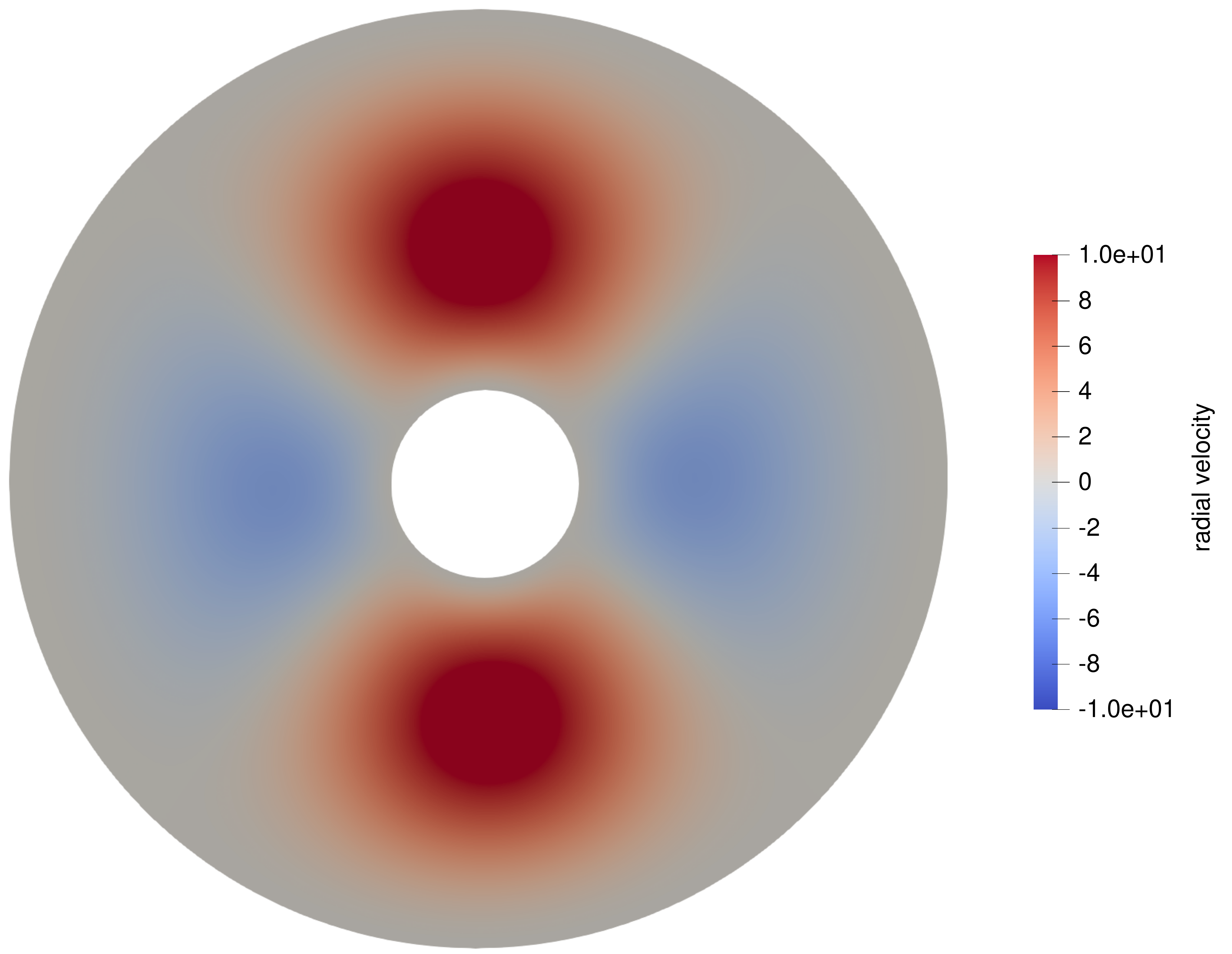} \\
(a) & (b) \\
& \\
& \\
\includegraphics[scale=0.14]{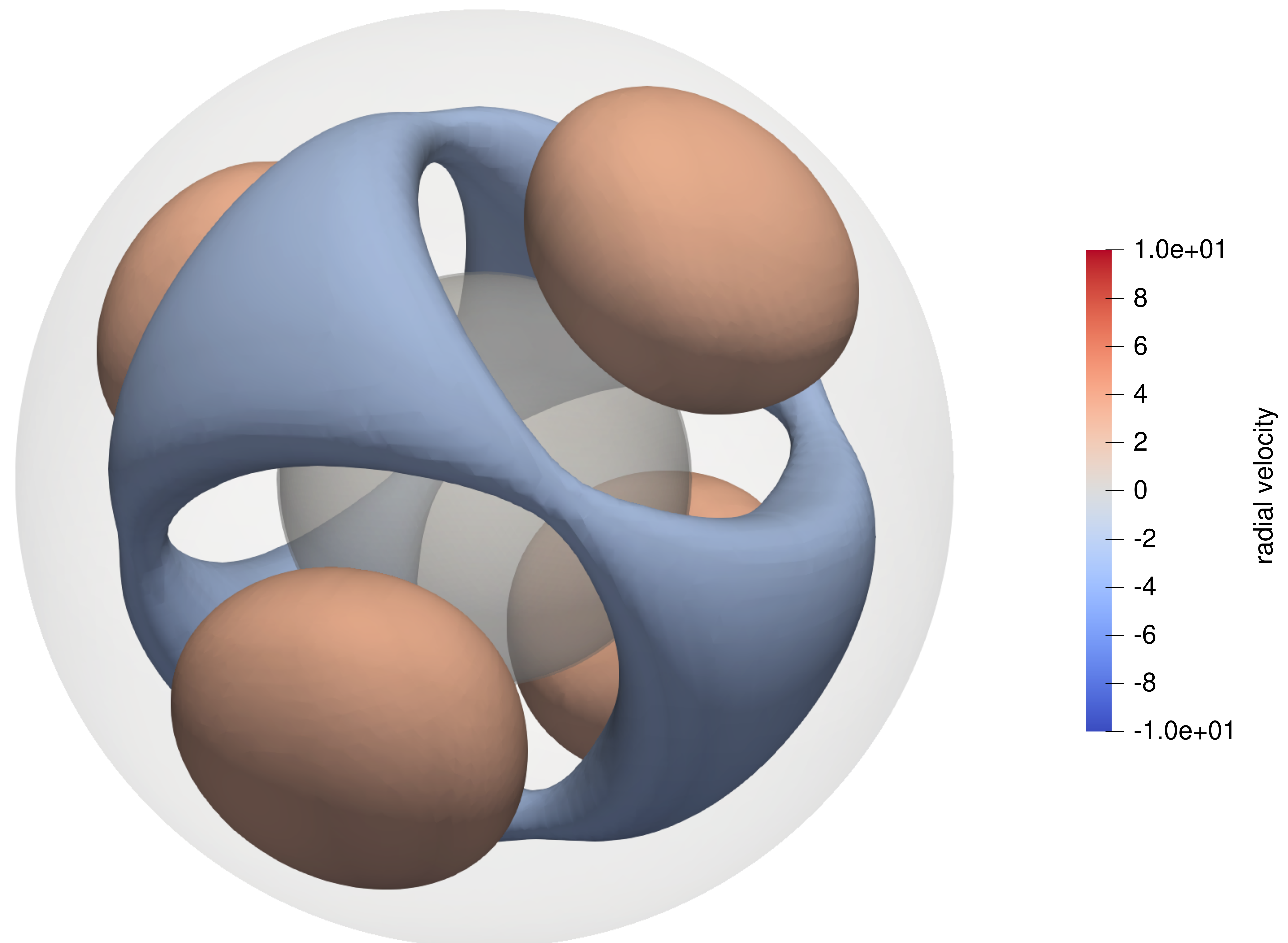} &
\includegraphics[scale=0.15]{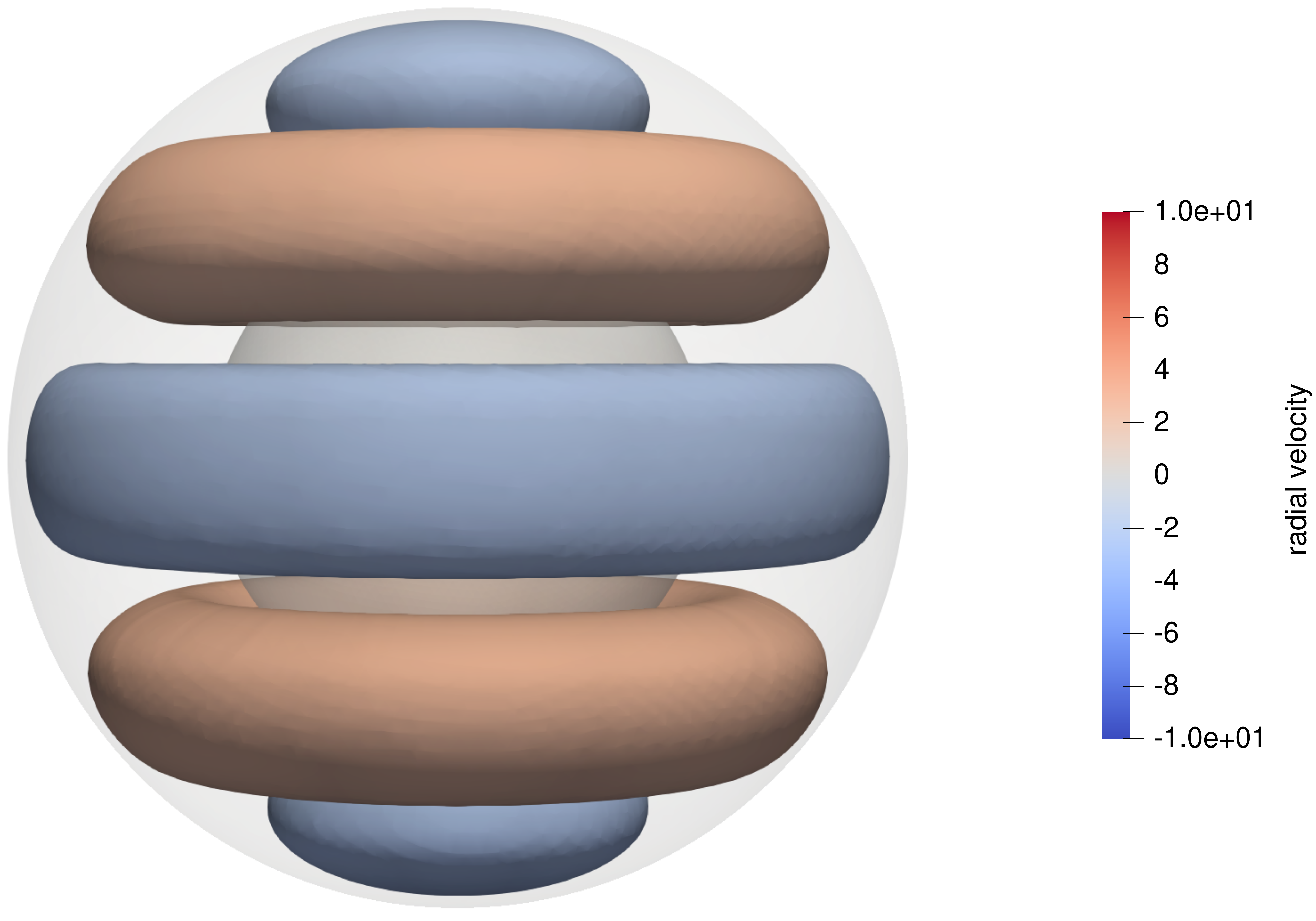} \\
(c) & (d) \\
& \\
& \\
\includegraphics[scale=0.13]{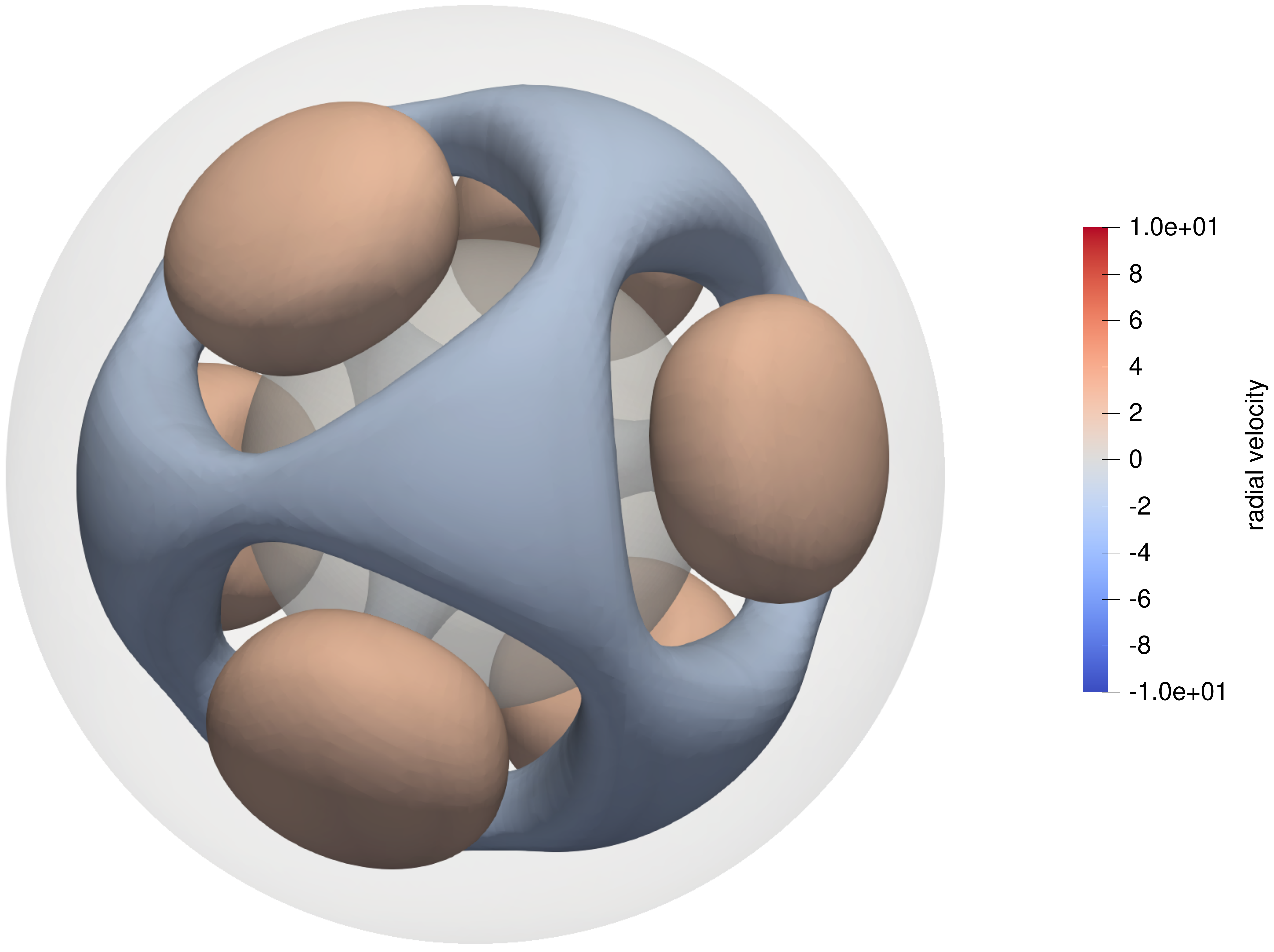} &
\includegraphics[scale=0.52]{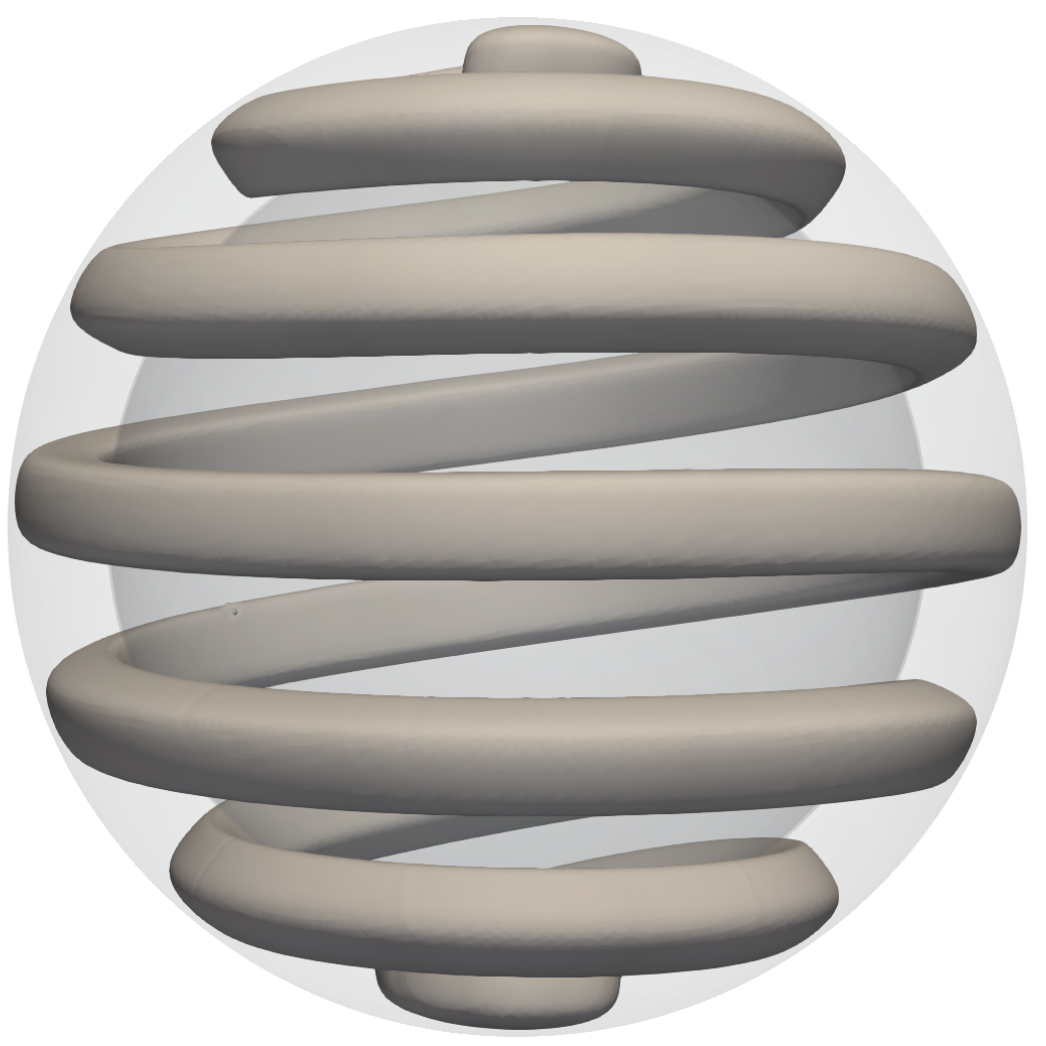} \\
(e) & (f) \\
\end{longtable}
\caption{ (a) and (b) Axisymmetric pattern (left) and it's meridional view (right) for $\bm{\eta = 0.2}$; (c) Tetrahedral convection pattern for $\bm{\eta = 0.4}$; (d) Axisymmetric pattern  for $\bm{\eta = 0.5}$; (e) Cubic pattern for $\bm{\eta = 0.5}$; (f) Spiral roll state for $\bm{\eta = 0.8}$ visualized by the contours of the radial velocity. Red color represents high or positive values and blue color indicates low or negative values.}
  \addtocounter{table}{-1}%
  \label{fig:convective patterns}
\end{figure}

Busse and Riahi \cite{busse1982patterns} reported that axisymmetric solution is preferred for the spherical shells which have $l_c = 2$, while the solution of tetrahedronal symmetry is preferred for $l_c = 3$. Recall that $l_c$ is the critical mode that triggers instability. Figure~\ref{fig:convective patterns}(a) shows the radial velocity isosurface for $\eta = 0.2$ with $l_c = 2$ signifying the axisymmetric pattern. The corresponding radial velocity contour on the meridional plane is shown in Fig.~\ref{fig:convective patterns}(b). Figure~\ref{fig:convective patterns}(c) shows convection with tetrahedral solution obtained for $\eta = 0.4$ with $l_c = 3$. 

For a spherical shell of $\eta = 0.5$, which has $l_c = 4$, we observe an axisymemtric pattern with $l=4$ and a cubic pattern, shown in Fig.~\ref{fig:convective patterns}(d) and Fig.~\ref{fig:convective patterns}(e), respectively, through the radial velocity isosurfaces. These patterns are in line with Arrial et al. \cite{arrial2014sensitivity}.

For $l_c = \mathcal{O}(10)$, Li et al. \cite{li2005multiplicity} and Itano et al. \cite{itano2015spiral} reported the formation of a single long spiral roll, extending from the north pole to the south pole, in the case of a moderately thin spherical shell as a stationary state. We qualitatively reproduce the spiral roll state for $\eta = 0.8$ as shown in the Fig. \ref{fig:convective patterns}(f). The spiral roll is formed by the superposition of spherical harmonics with degrees $l = 13$ and $14$, with $l_c = 13$ being the critical mode for $\eta = 0.8$. Note that the initial random fields for this case of $\eta = 0.8$ are free from any symmetry.

%

\FloatBarrier
\FloatBarrier

\subsection{Rayleigh-B\'enard convection}
\label{subsection: Rayleigh Benard convection}

For the test cases in this subsection, we adopt the formulation of Gastine et al. \cite{gastine2015turbulent} along with the DEC-FD discretization methodology and compare the output parameters against \cite{gastine2015turbulent}. The spherical shell is now subject to basal heating with gravity profile $\bm{g} = -\left( R_0/r \right)^2 \bm{e_r}$, where $g(r)=\left( R_0/r \right)^2$. Here, the Boussinesq equations are non-dimensionalized using the thickness of the spherical shell $d = R_1-R_0$ as the reference length scale, the viscous dissipation time $d^2/\nu$ as the time scale, and the temperature difference $\Delta T = T_1 - T_0$ as the temperature scale, where $T_1$ and $T_0$ are the temperatures at the top and bottom spherical surfaces. The non-dimensionalized radial conductive temperature profile with isothermal boundaries is

\begin{equation}
T_c(r) = \frac{\eta}{\left( 1-\eta \right)^2}\frac{1}{r} - \frac{\eta}{\left( 1-\eta \right)}, \ \ \ T_c(r = R_0) = 1, \ \ \ T_c(r = R_1) = 0,
\end{equation}

\noindent which results in $f(r) = \eta / \left[ r^2 \left ( 1- \eta \right)^2 \right]$. The gravity is nondimensionalized using its reference value $g_0$ at the outer boundary. The Rayleigh number $Ra$ then turns out to be

\begin{equation}
Ra = \frac{\alpha g_0 \Delta T d^3}{\nu \kappa}.
\label{eq: non-dimensional control paramters}
\end{equation}

\noindent Based on this non-dimensionalization and preceding definitions for $g(r)$ and $f(r)$, the coefficients in Eqs. (\ref{eq: non-dimensional continuity}) $-$ (\ref{eq: non-dimensional energy equation}) turn out to be 

\begin{equation}
\zeta(r) = \frac{Ra}{Pr}g, ~~~ \psi = 1, ~~~ \chi(r) = \frac{\eta}{\left (1- \eta \right)^2} \frac{1}{r^2}, ~~~  \xi = \frac{1}{Pr}.
\end{equation}

\textcolor{black}{To define the output parameters, we employ the following notation;} (1) Time average denoted by $\overline{\cdots}$; (2) Spatial average over the entire volume of the spherical shell represented by $\langle\cdots\rangle$; (3) Spatial average over a spherical surface indicated by $\langle\cdots\rangle_s$; to obtain the averaging procedures:

\begin{equation}
\overline{f} = \frac{1}{\tau}\int_{t_o}^{t_o + \tau} f \  \text{d}t, \ \ \ 
\langle f \rangle = \frac{1}{V}\int_V f(r,\theta,\phi) \  \text{d}V, \ \ \ 
\langle f \rangle_s = \frac{1}{a}\int_a f (r,\theta,\phi) \ \text{d}a,
\label{eq: non-dimensional control paramters}
\end{equation}

\noindent where $\tau$ is the time averaging interval, $V$ is the volume of the spherical shell, and $a$ is the area of the spherical surface. The heat transport is characterized by Nusselt number $Nu$, which is the ratio of total heat transfer (convective and conductive) to the conductive heat transfer. The Reynolds number $Re$ is given by the root-mean-square flow velocity. The Nusselt number and Reynolds number are defined as

\begin{equation}
Nu = -\eta \frac{\text{d}v}{\text{d}r} \left( r = R_0 \right), \ \ \ Re = \overline{\sqrt{\langle u_r^2 + u_\perp^2 \rangle}},
\label{eq: non-dimensional control paramters}
\end{equation}

\noindent where $v(r)= \overline{ \langle T \rangle_s}$ is the time averaged and surface averaged radial temperature profile. 
We quantify $Nu$ and $Re$ for $\eta = 0.6$, $Pr = 1$, and $1.5\times10^3 \leq Ra \leq 3\times10^4$ in the Table \ref{table: RBC quantification}, and a maximum error of $3\%$ in comparison with \cite{gastine2015turbulent} is observed.


\begin{table}[htp!]
\caption{Prediction of Nusselt number and Reynolds number for the numerical simulations of Rayleigh-B\'enard convection for $\bm{Pr = 1}$ and $\bm{\eta = 0.6}$. }
\begin{center}
\setlength{\tabcolsep}{25pt}
\renewcommand{\arraystretch}{1.2}
\begin{tabular}{c c c c c}
\hline
\hline
 & \multicolumn{2}{c}{Present study} & \multicolumn{2}{c}{Gastine et al. \cite{gastine2015turbulent}}\\
$Ra$ & $Nu$ & $Re$ & $Nu$ & $Re$ \\
\hline 
1.5 $\times 10^3$ & 1.33 & 4.41 & 1.33 & 4.4 \\
3 $\times 10^3$ & 1.84 & 9.84 & 1.80 & 9.6 \\
5 $\times 10^3$ & 2.08 & 14.09 & 2.13 & 14.4 \\
1 $\times 10^4$ & 2.53 & 23.74 & 2.51 & 23.3 \\
2 $\times 10^4$ & 3.09 & 35.75 & 3.05 & 35.0 \\
3 $\times 10^4$ & 3.54 & 43.64 & 3.40 & 44.0 \\
\hline
\end{tabular}
\end{center}
\label{table: RBC quantification}
\end{table}%
\FloatBarrier

\section{Conclusion}
\label{section: Conclusion}

In this work, we propose a novel hybrid discrete exterior calculus and finite difference (DEC-FD) method to simulate fully three-dimensional Boussinesq convection in spherical shells for a wide range of aspect ratios. The grid used in this method is free of the problems like coordinate singularity, grid non-convergence near the poles, and overlap regions, which are often encountered in the Yin-Yang and cubed-sphere grids. We derive the operator split governing equations and replace the surface operators with DEC operators, and \textcolor{black}{approximate the radial operators using FD operators}. The new DEC-FD formulation is verified using a series of numerical examples that are simulated using an in-house parallel code developed based on the PETSc framework. We observe an excellent agreement of the critical Rayleigh number estimated numerically against the analytical results derived by Chandrasekhar \cite{chandrasekhar2013hydrodynamic}, with a maximum error being $9\%$. For Rayleigh-B\'enard convection, the output parameters, the Nusselt and the Reynolds numbers, are predicted with an error less than $3\%$ in comparison to Gastine et al. \cite{gastine2015turbulent}. In addition, we reproduce a robust, long spiral roll, covering all the spherical surface, for a moderately thin spherical shell, which is in line with results reported by Li et al. \cite{li2005multiplicity} and Itano et al. \cite{itano2015spiral}.

We expect to see future extensions of this work implementing a tuned centered finite difference scheme in the radial direction, which will improve the resolution properties of the overall scheme. Future development of this code will deal with iterative solvers and mesh distribution for massively parallel computations to explore extreme parameter regimes ($Ra \geq 10^9$).
\section*{Acknowledgments}
\textcolor{black}{BM and PJ are dismayed to note the untimely demise on July 24, 2022 of their coauthor, Ravi Samtaney, during the revision of this manuscript. }This publication is based upon work supported by the King Abdullah University of Science and Technology (KAUST) Office of Sponsored Research (OSR) under Award URF/1/4342-01. For computer time, this research used the Cray XC40, Shaheen II, of the Supercomputing Laboratory at King Abdullah University of Science \& Technology (KAUST) in Thuwal, Saudi Arabia.

\bibliography{reference.bib}
\bibliographystyle{elsarticle-num.bst}

\FloatBarrier

\FloatBarrier

\end{document}